\documentclass[preprint]{aastex}

\slugcomment{April 22 2016 - Accepted by ApJ}

\shortauthors{Millan-Gabet et al.}

\begin{document}


\title{Confronting Standard Models of Proto--Planetary Disks With New Mid--Infrared Sizes from the Keck Interferometer}

\author{Rafael Millan-Gabet}
\affil{California Institute of Technology, NASA Exoplanet Science
  Institute, Pasadena, CA 91125, USA}
\email{R.Millan-Gabet@caltech.edu}

\author{Xiao Che}
\affil{University of Michigan Astronomy Department, 1085 S. University Ave. 303B West Hall University of Michigan, Ann Arbor, MI 48109-1107, USA}

\author{John D. Monnier}
\affil{University of Michigan Astronomy Department, 1085 S. University Ave. 303B West Hall University of Michigan, Ann Arbor, MI 48109-1107, USA}

\author{Michael L. Sitko}
\affil{Department of Physics, University of Cincinnati, Cincinnati OH 45221, USA}
\affil{Center for Extrasolar Planetary Systems, Space Science Institute, Boulder, CO 80301}

\author{Ray W. Russell}
\affil{The Aerospace Corporation, Los Angeles, CA 90009, USA}

\author{Carol A.Grady}
\affil{Eureka Scientific, 2452 Delmer, Suite 100, Oakland, CA 96002, USA}

\author{Amanda N. Day}
\affil{Department of Physics, University of Cincinnati, Cincinnati OH 45221, USA}

\author{R. B. Perry}
\affil{NASA Langley Research Center, MS 160, Hampton, VA 23681}

\author{Tim J. Harries}
\affil{Department of Physics and Astronomy, University of Exeter, Stocker Road, Exeter EX4 4QL, UK}

\author{Alicia N. Aarnio}
\affil{University of Michigan Astronomy Department, 1085 S. University Ave. 303B West Hall University of Michigan, Ann Arbor, MI 48109-1107, USA}

\author{Mark M. Colavita}
\affil{Jet Propulsion Laboratory, California Institute of Technology, 4800 Oak Grove Drive, Pasadena, CA 91109, USA}

\author{Peter L. Wizinowich}
\affil{Keck Observatory, 65-1120 Mamalahoa Hwy, Kamuela, HI 96743, USA}

\author{Sam Ragland}
\affil{Keck Observatory, 65-1120 Mamalahoa Hwy, Kamuela, HI 96743, USA}

\and

\author{Julien Woillez\altaffilmark{1}}
\affil{Keck Observatory, 65-1120 Mamalahoa Hwy, Kamuela, HI 96743, USA}

\altaffiltext{1}{European Southern Observatory, Karl-Schwarzschild-Strasse 2, Garching D-85748, Germany}

\begin{abstract}

We present near and mid--infrared interferometric observations made with the Keck Interferometer Nuller and near--contemporaneous spectro--photometry from the IRTF of 11 well known young stellar objects, several observed for the first time in these spectral and spatial resolution regimes.  With AU--level spatial resolution, we first establish characteristic sizes of the infrared emission using a simple geometrical model consisting of a hot inner rim and mid--infrared disk emission.   We find a high degree of correlation between the stellar luminosity and the mid--infrared disk sizes after using near--infrared data to remove the contribution from the inner rim.  We then use a semi--analytical physical model to also find that the very widely used ``star~$+$~inner~dust~rim$+$~flared~disk'' class of models strongly fails to reproduce the SED and spatially--resolved mid--infrared data simultaneously; specifically a more compact source of mid--infrared emission is required than results from the standard flared disk model. We explore the viability of a modification to the model whereby a second dust rim containing smaller dust grains is added, and find that the two--rim model leads to significantly improved fits in most cases. This complexity is largely missed when carrying out SED modelling alone, although detailed silicate feature fitting by \citet{mcc2013} recently came to a similar conclusion.  As has been suggested recently by \citet{menu2015}, the difficulty in predicting mid--infrared sizes from the SED alone might hint at ``transition disk''--like gaps in the inner AU; however, the relatively high correlation found in our mid--infrared disk size vs. stellar luminosity relation favors layered disk morphologies and points to missing disk model ingredients instead.

\end{abstract}


\section{Introduction}

The gas and dust disks around young stars play an important role in the formation and evolution of stars and planetary systems.  A protostellar object grows as it accretes matter from its circumstellar disk.  At the same time, the physical conditions in the disks constitute the initial conditions for planet formation \citep{wil2011}. It is therefore important to know the disk structure and composition as a function of stellocentric radius and vertical height, density and temperature  profiles of each disk component, and how these properties evolve with time, in order to improve our theoretical understanding of the planet formation processes \citep{bod2002,blum2008}. Direct observational constraints are however difficult to obtain, due to angular resolution limitations inherent to standard imaging techniques, as we now illustrate. 

Generally speaking, mid--infrared (MIR) wavelengths probe disk emission from ``intermediate'' radial locations, between the innermost disk regions bright in the near--infrared (NIR) and the outer disk emitting at (sub)--mm wavelengths and also visible in scattered light images \citep[see e.g. Figure~1 in][]{dul2010}. For an A0 star, for example, \citet{boek2005} place 90\% of the MIR disk emission between 0.5 -- 30~AU. Therefore, this wavelength regime is interesting as it probes the spatial scales where planets form and reside. At typical distances to star forming regions however ($d > 100$~pc), these spatial scales ($\lesssim 300$~mas) are hardly resolved using conventional telescopes. For this reason progress has relied mostly on interpreting spectral energy distributions (SEDs), which have inherent degeneracies (most notably between disk temperature and dust properties) and therefore necessarily rely on disk models for which even the most basic aspects pertaining to the innermost regions have not been solidly established.

Long baseline interferometers operating at MIR wavelengths can spatially resolve the relevant spatial scales, and provide much needed new model constraints. Previous surveys have focused on establishing the characteristic MIR sizes of a relatively small number of T~Tauri and Herbig~Ae/Be objects \citep{lein2004}, including results at lower spatial resolution using specialized interferometric techniques on single large telescopes \citep{hinz2001,liu2005,liu2007,jdm2009}. First steps have also been taken in exploring the dust mineralogy and showing that the distribution of dust species is not homogeneous in the disk \citep{boek2004} and comparing with parametrized disk models \citep{fed2008,sche2009}. Modelling the MIR emission in detail however is notoriously complicated, because it contains contributions from several disk regions, as well as fundamental uncertainties about whether or not the relevant disk regions are completely or partially shadowed. This is in contrast with the modelling of the NIR emission, which is almost completely dominated by a single disk component -- namely the inner dust rim \citep[there are also smaller contributions from inner gas and outer dust envelope][and references therein]{dul2010}.  Indeed, a handful of single--object studies using specific detailed disk models have provided valuable insights, but also illustrate the difficulty of the problem \citep{kraus2008,sche2008,folco2009,ratz2009,ben2010,rag2012,sche2013,gab2013}. Most recently, \citet{menu2015} present the results of a survey of 41 Herbig Ae/Be objects with the MIDI instrument at the Very Large Telescope Interferometer. They find intrinsic morphological disk diversity or evolutionary diversity, and evidence for flat disks (group II) having gaps, with implications for the evolutionary sequence and possible role of planet formation in producing the observed types of disks (flat with or without gaps, and flared/gapped -- i.e. transitional).

In this paper we present new spatially resolved observations using the Keck Nuller Interferometer (KIN) of the NIR and MIR brightness for 11 well known young stellar objects (YSOs), as well as near--contemporaneous spectro--photometric data obtained at the NASA Infrared Telescope Facility (IRTF). We do not attempt to constrain the parameters of a specific detailed physical model, because the amount of data available would not permit us to resolve the many model parameter degeneracies, and would result in a very limited gain in knowledge, especially considering that those detailed physical models are themselves still largely unproven. Rather, our approach is to use simple and general model prescriptions that still reflect the most salient physical processes, in order to establish the basic features of the infrared brightness, test current paradigms, and suggest directions to improve the models.

\section{The Sample}

Our sample consists of 11 targets selected to have strong infrared excess flux over the stellar photospheres. They represent four different YSO types: 3 T~Tauri, 4 Herbig~Ae, 3 Herbig~Be, and 1 FU~Ori object. Their basic properties, and the parameters needed for the modeling performed in the sections that follow are shown in Table~\ref{starlist}. All the targets are well known young circumstellar disk objects, and the disk properties adopted, also inputs to the modeling, are listed in Table~\ref{disk properties}.

\section{Observations and Data Reduction}

Observations were made using the Keck Interferometer \citep{col2013} in its nuller mode \citep{col2009}, and at the NASA Infrared Telescope Facility (IRTF) over the time period 2009--2010 -- see the observing log in Table~\ref{obslog}. 

\subsection{Keck Nulling Interferometry}

The Keck Interferometer Nuller \citep[KIN,][]{col2009} operates in N--band ($8.0 - 13.0 \, \mu$m, dispersed over 10 spectral pixels)  and combines the light from the two Keck telescopes as an interferometer with a physical baseline length $B \sim 85$~m. The KIN produces a dark fringe through the phase center (``Nulling''). The adjacent bright fringe (through which flux is transmitted), projects onto the sky at an angular separation $ \lambda/2B = 10$~mas, or 1.4~AU at the median distance to the stars in our sample (140~pc), and for $\lambda = 8.5 \, \mu$m (the effective wavelength of the KIN bandpass). Thus, the instrument is sensitive to MIR circumstellar emission as close to the central star as these spatial scales (i.e. ``inner working angle''). For further descriptions of the KIN observables, see \citet{mil2011}, \citet{ser2012}, or \citet{men2014}.

The KIN also uses a standard Michelson interferometer operating in K--band ($2.0 - 2.4 \, \mu$m, dispersed over 5 spectral pixels), as a fringe tracker in order to stabilize the MIR nulls in the presence of optical path fluctuations induced by the turbulent Earth's atmosphere. In this paper we also use these NIR interferometric data, in order to probe circumstellar emission from hotter disk regions located closer to the central star. For the physical baseline length, the fringe spacing at $2.2 \, \mu$m is 5.3~mas, or 0.8~AU at the median distance to our sample.

The MIR nulls and NIR visibility data provided by the KI pipeline were calibrated using their {\em Calib} package\footnote{http://nexsci.caltech.edu/software/KISupport/} . Following standard practice, in order to measure the instrument's transfer function and account for it in the data calibration process, observations of targets of interest were interleaved with observations of calibrator stars of known angular diameters (see Table \ref{obslog}). For ease of comparison of the MIR and NIR data, the calibrated nulls ($n$) were converted to visibilities using the relation $V = (1 - n)/(1 + n) $ -- an appropriate approximation given that the MIR emission from our sources appears essentially unresolved to the 4~m baseline of the KIN cross--combiner \citep[see][]{col2009}. A salient aspect of the MIR spatially resolved measurements presented in this paper is that due to the nulling mode, the precision of the calibrated MIR visibilities is substantially higher than can be achieved with standard MIR interferometers from the ground \citep{col2009,col2010}. Our typical uncertainties are $\sigma_{n} = 0.005 - 0.01$, depending on observing conditions and on the spatial extent of the object in the NIR fringe tracking channel; which corresponds to MIR visibility uncertainties $1 - 2$\% for an unresolved object. 

\subsection{IRTF Spectrophotometry}

For most of the KIN objects and epochs, we also obtained new NIR and MIR spectrophotometric data at the IRTF. Best attempts were made to schedule the IRTF observations as near--contemporaneously with the KIN observations as possible, in practice resulting in time lags ranging from a few days to two months, one month being typical (see Table~\ref{obslog}). This is important because temporal variations in the star/disk flux ratios are known to be common among YSOs \citep{sit2008}, and accurate relative fluxes are needed input to the modelling of the interferometric visibilities. Within the time interval between the KIN and spectrophotometric data, we assume that the disk morphology and star/disk flux ratios remain constant.

We obtained NIR spectra using the SpeX spectrograph \citep{ray2003}. The spectra were recorded using the echelle grating in both short--wavelength mode (SXD, $0.8-2.4 \, \mu$m) and long wavelength mode (LXD, $2.3-5.4 \, \mu$m) using a 0.8 arcsec slit. The spectra were corrected for telluric extinction and flux calibrated against a number of A0~V calibrator stars, using the Spextool data reduction package \citep{cus2004,vac2003}. 

In addition to the 0.8~arcsec---slit spectra, for all but v1295~Aql and v1057~Cyg we also recorded data with the SpeX prism disperser and a wide 3.0 arcsec slit, which allows us to retrieve the absolute flux levels when the sky transparency is good and the seeing is 1~arcsec or better. This condition was met for DG~Tau, RY~Tau, MWC~480, and AB~Aur, and confirmed using the BASS data, obtained a month (DG~Tau and RY~Tau) or 2 days (MWC~480 and AB~Aur) in time from the SpeX observations. For v1295~Aql and v1057~Cyg we normalized the SpeX levels using the BASS observations alone, which were  obtained within a week of the SpeX observations. For MWC 275, the seeing was 1.4 sec, but the Prism and BASS yielded identical scaling factors for the SXD+LXD spectra.

MIR spectra were obtained with The Aerospace Corporation's Broad-band Array Spectrograph System (BASS). BASS uses a cold beamsplitter to separate the light into two separate wavelength regimes. The short--wavelength beam includes light from $2.9-6 \mu$m, while the long--wavelength beam covers $6-13.5 \, \mu$m. Each beam is dispersed onto a 58--element Blocked Impurity Band (BIB) linear array, thus allowing for simultaneous coverage of the spectrum from $2.9-13.5 \, \mu$m. The spectral resolution $R = \lambda$/$\Delta\lambda$ is wavelength--dependent, ranging from about 30 to 125 over each of the two wavelength regions \citep{hac1990}. In some cases where the wide--slit SpeX Prism observations were not available, BASS spectrophotometry that overlapped the SpeX data were used to provide absolute flux levels of the SpeX spectra.

In order to construct complete SEDs for each object, additional infrared photometry from 2MASS, Spitzer and the literature have been included as needed in order to fill in wavelengths gaps in either the Spex or BASS data. The UBVRI data are primarily from the EXPORT project \citep{oud2001} or from the survey of HAeBe stars published by \citet{win2001}. 


\section{Modeling and Results}

\subsection{Stellar Photosphere}

In order to study the disk emission, it is necessary to estimate the stellar contribution to the observed SEDs. It is reasonable to assume that shorter wavelength fluxes are dominated by the stellar photosphere, because the circumstellar disks are much cooler. Therefore, we fit a stellar model to the UBVRI SED data, and extrapolate the modeled stellar spectra to the longer wavelengths at which KIN operates. 

We use Kurucz models for the stellar photospheres \citep{kur1979}. The stellar metallicity is assumed to be solar, and stellar masses and distances are fixed to the values listed in Table~\ref{starlist}. The parameters we fit are: stellar surface effective temperature ($T_{\star}$), radius ($R_{\star}$) and reddening coefficient (including circumstellar material). The best--fit results are shown in Table~\ref{stellarparam}. Our values are consistent with previous SED--based results in the literature. When modelling the disk emission, as described in the following sections, the stellar contributions to the SED are fixed to these best--fit results. 



\subsection{Geometric Disk Model} 
\label{sec:geom}

\subsubsection{Model and Fitting Procedure}
\label{sec:desc}

We begin by using a geometric disk model in order to establish the emission size scales. The objects are represented as a linear combination of the three components expected to dominate the emission: the star, the inner dust rim, and the extended disk behind it. 

\begin{figure} 
\begin{center} { \includegraphics[width = 6.5in]{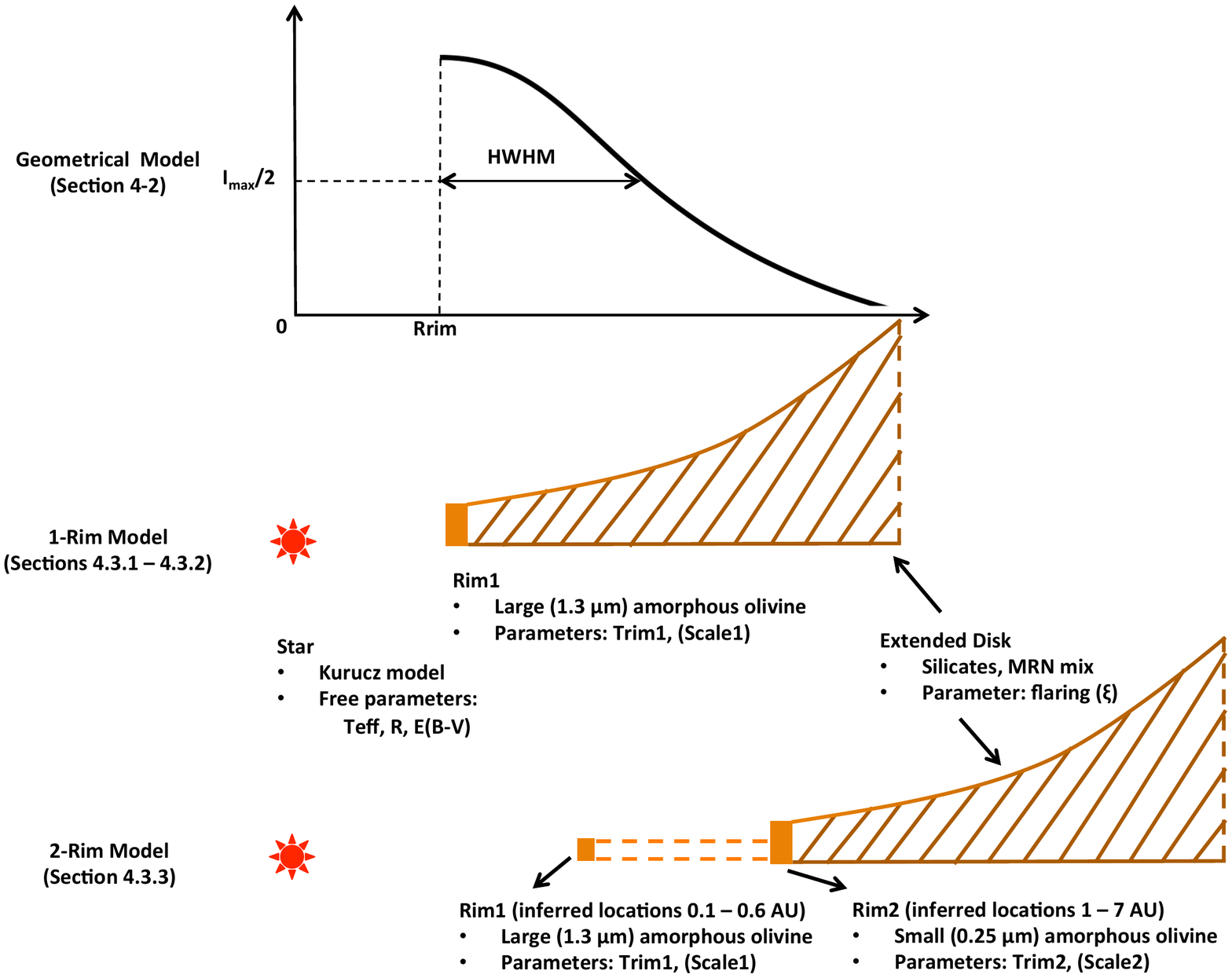} } 
\hphantom{.....} 
\caption{Schematic representation of the inner disk models used in this paper and the relevant parameters in each case. We note that in the 1--rim and 2--rim model schematics the intent is to only indicate the average locations of the rims, with no physical meaning to the extent of the region between them (see text).
\label{fig: schematic disk} }
\end{center} 
\end{figure}

The star is modelled as an unresolved point source, which is appropriate given their angular diameters (all smaller than 0.2~mas) and angular resolution of the KIN (5~mas fringe spacing at even the shortest 2.2~$\mu$m wavelengths in these observations). The inner dust rim is represented by a ring of linear radius $R_{rim}$, infinitely thin in the radial direction, and emitting as a blackbody at temperature $T_{rim}$. The emission from the extended disk is represented by a  two--dimensional Gaussian brightness with a central clearing of radius equal to the inner rim radius, we quantify the size scale of this component by its half--width at half--maximum ($\mbox{HWHM}_{Disk}$, see Figure~\ref{fig: schematic disk}). The inclination and position angle of both the rim and extended disk are assumed to be the same as those observed via millimeter interferometry of the outer disk  (given in Table \ref{disk properties}). The fitting process is divided into two steps, as follows. 

First, the temperature and size of the inner rim ($T_{rim}$, $R_{rim}$) are determined
from the NIR SED and K--band visibilities, ignoring the extended disk component since it contributes negligible flux at NIR wavelengths (in practice, we limit the SED fits to the 1--5~$\mu$m wavelength region in order to best realize this assumption).  The inner rim temperature is obtained by fitting the NIR SED (its shape constrains this parameter very well). The rim radius is then obtained by numerically solving the equation for the K--band visibilities:

\begin{equation} 
V(\lambda, u, v) = \left( \frac{F_{\star}}{F}  \right)_{\lambda} + \left( \frac{F_{rim}}{F} \right)_{\lambda} \cdot V_{rim}(\lambda,u,v)  =\left( \frac{F_{\star}}{F}  \right)_{\lambda} + \left( \frac{F_{rim}}{F} \right)_{\lambda} \cdot J_{0}(2 \pi \rho b)
\end{equation}

where  $V$ is the observed visibility amplitude at each of the 5 wavelength bins sampled within the K--band,  $F$ the total flux, $F_{\star}$ the stellar flux,  $F_{rim}$ the rim flux, $\rho = R_{rim}/d$ is the angular radius of the rim, and $b$ is the projected baseline ($b = \sqrt{u^2+v^2}/\lambda$) taking into account the inclination and  orientation of the rim on the sky. The fractional fluxes are obtained by SED decomposition using the stellar fit described  above. Therefore the only unknown is the radius of the rim $\rho$. The Bessel function ($J_0$) in the equation above is not bijective; here we consider only numerical solutions in the main lobe of the visibility function, i.e. we adopt the {\em smallest} rim size consistent with the data.

Next, we determine the characteristic size of the extended disk by fitting to the N--band visibilities. This time the star is ignored because it contributes negligible flux in N--band. Therefore, the spatial model consists of the inner rim (barely resolved at MIR wavelengths -- see Table~\ref{geom}) and the extended disk component. Similarly to the previous step, the fractional fluxes in each component at the 10 N--band wavelength bins are obtained via SED decomposition, using the parameters for the blackbody ring representing the inner rim from the previous step. Therefore, the $\mbox{HWHM}$ of the truncated Gaussian brightness representing the extended disk is the only free parameter. In practice, the fitting is performed by generating an image of this model and the visibilities are extracted via Fourier transformation. 

\subsubsection{NIR and MIR Characteristic Sizes}
\label{sec:results}

Figure~\ref{fig: geom} shows the SED data, visibility data, and fitted sizes (i.e. radii given by $R_{rim}$ in the NIR or $R_{rim}+\mbox{HWHM}_{Disk}$ in the MIR) as a function of wavelength within each of those bandpasses. Table~\ref{geom} shows the best--fit parameters for each object; where the rim and extended disk radii have been averaged over the spectral bins in the NIR and MIR bandpasses respectively (for the propagation of errors, we assume that the NIR spectral bins are uncorrelated, and that the MIR spectral bins are fully correlated, following \citet{men2014}). We note that the uncertainties in the characteristic sizes in Table~\ref{geom} do not include systematic uncertainties due uncertainties in (a) the fractional fluxes derived via SED decomposition (for reference, a $\sim 10 \%$ effect given our photometric errors and values of the $J_{0}$ term in Eq.~1 typical of our sample), or (b) distance (a $25 \%$ effect given the same level of distance uncertainties for our sample).

We obtain best--fit values for the rim temperatures and radii that are in agreement with expected dust sublimation values, as was previously found \citep[see e.g.][and references therein]{dul2010}. The MIR characteristic sizes range from 1.2~AU to 6.7~AU, with median precision of 3\%.

We note that (as can be seen in Figure~\ref{fig: geom}) for AB~Aur, as well as for RY~Tau and MWC~758 at some of the MIR wavelengths, there is no Gaussian $\mbox{HWHM}$ solution. This is because for those cases the coherent MIR flux (MIR visibility times the total flux, solid orange line in the SED panels) is lower than the rim flux, and therefore there is no mathematical solution for the Gaussian component, given that as noted above the rims are nearly unresolved at MIR wavelengths. In other words, our procedure for this simple geometrical model places too much MIR coherent flux in the rim. In Section~\ref{sec:physmodel} we consider more physical models which allow for a more extended MIR brightness for these sources.

\subsubsection{The MIR Size -- Stellar Luminosity Relation}

Studying how the characteristic sizes relate to the stellar properties can reveal clues about the dominant emission processes at play in a given wavelength regime. In Figure~\ref{fig: size-l} we explore how the MIR characteristic sizes measured above ($\overline{R}_{rim} + \overline{HWHM}_{Disk}$) relate to the stellar luminosity ($L_\star$). The index number in the plot identifies each object as in Table~\ref{geom} (AB~Aur is missing, because the geometrical model has no MIR size solution for this object, as discussed above). The dashed lines represent the equilibrium location of gray dust at the indicated temperatures, following the definition of \citet{m&m2002}. 

We confirm earlier findings that the MIR sizes generally scale with stellar luminosity \citep{jdm2009,menu2015}. However, we find a better correlation than found by these previous authors. Formally, we find a correlation of 0.9 with a low p--value (0.001) indicating that the null hypothesis (no correlation) is rejected. Alternatively, a bootstrap analysis gives a 5$\sigma$ significance to the measured slope of the MIR~size vs. $L_\star$ diagram ($\mbox{slope} = 0.19 \pm 0.04$).  

Most likely, the reason for the higher level of correlation is that our choice of ``MIR~size'' effectively removes the rim emission, so that the remaining MIR size correlates better with stellar luminosity. Our two--step procedure is indeed very different from e.g. the one--component Gaussian model of \citet{jdm2009} or the half--light at half--radius measure of a T--power law disk of \citet{menu2015}. The lower scatter in our relation may also be, at least in part, the result of using the known inclination of the (outer) disk for each object (Table~\ref{disk properties}), rather than uniformly assuming a face--on geometry.






\subsection{Semi--analytical Model: Flared Disk with Inner Dust Rim}
\label{sec:physmodel}

We now turn our attention to determining how the new MIR interferometer data compares with predictions from a physical model that encapsulates current paradigms -- namely a flared--disk including a ``puffed--up'' inner dust rim \citep[see e.g.][and references therein]{ dul2010}. We use our own semi--analytical implementation of this model, so that we can modify it, which we will show may be necessary. Our semi--analytical model follows \citet{eis2004} but with an inner rim following \citet{dal2004} and \citet{ise2005}. 
As in the previous section, the inclination and position angle of the rim and flared disk are assumed to be the same, and we use the values inferred from millimeter interferometry of the outer disk (Table~\ref{disk properties}).


For simplicity, dust grains in the rim are assumed to be a single species, namely amorphous olivine MgFeSiO(4), commonly found in circumstellar disks \citep{dor1995,sar2009} with close to cosmic Mg--to--Fe ratio \citep[e.g.][]{sno1995}. The optical constants for this species are from \citet{jae1994} and \citet{dor1995}. The opacities are computed using Mie theory. Since the rim is hot, only large grains can survive; here we assume a single size of $1.3\,\mu$m \citep{tan2008b}.

With the rim grain properties fixed as discussed above, the rim radius is determined by the sublimation temperature \citep{dal2004, ise2005}. Thus, the model for the rim component has two free parameters: a scale parameter related to the angular size of the projected rim surface, used to match the NIR fluxes, and the dust sublimation temperature ($T_{rim}$).


Dust grains in the flaring disk component behind the dust rim are assumed to be silicates with optical properties as in \cite{lao1993}. Our calculations showed that the dust grain size upper cutoff is not important for our results in the MIR wavelength range, therefore we used a standard MRN distribution \citep{mat1977} with grain sizes following a power law with index -3.5, and minimum/maximum sizes of 0.005/0.25~$\mu$m respectively.   

The mass and outer radius of the flared disk component are fixed to the values in Table~\ref{disk properties}. We note that the outer radius has negligible effect on the predicted SED or the interferometric data at MIR and shorter wavelengths, because the outer disk regions contribute little NIR or MIR emission. The surface density distribution is assumed to follow a power law with index of -1.5 \citep{chi1997}. The only free parameter is the flaring index\footnote{defined such that the disk height above the mid--plane increases with radial distance from the star as $h/r \propto r^{\xi}$.} $\xi$, which determines how much stellar emission the extended disk can intercept (i.e. the larger the flaring index, the hotter the extended disk is). Validation of our semi--analytical implementation of the flared disk model against benchmark radiative transfer codes is presented in Appendix~A.

We note that for the purposes of this exercise, we do not use the NIR interferometer data. This is because \citet{tan2008b} showed that in order to explain the shape of the NIR visibility curves past the first lobe, a relatively smooth NIR brightness was required (i.e. inconsistent with the abrupt edge in the NIR brightness that results from models devoid of emission inside the inner dust rim). They argued that the most likely origin of the extra NIR emission is hot gas interior to the dust sublimation radius, a component clearly not included in the model just described.

Finally, we note that for 4 of the 11 objects: DG~Tau, MWC~1080, v1057~Cyg, and v1685~Cyg, our model has no hope of reproducing the detailed SED, because for those objects no silicate emission feature is observed. Possible reasons are: (a) the disks contain only Carbon grains (a radical possibility), or (b) the MIR excess arises in an optically thick envelope of large grains, or (c) large gaps exist in the disk region normally responsible for silicate emission \citep{maa2013}. Indeed these 4 objects are known to be very active and/or embedded, such that our model clearly does not apply, and the tailored models that would be required are outside the scope of this paper. However, we choose to keep those four objects in the rest of our analysis, because it is still valuable to examine how the model fares in reproducing not the details but the general features of the data, namely the infrared excess and MIR visibilities {\it levels}. A schematic sketch of the model and parameters is shown in Figure~\ref{fig: schematic disk}. 





\subsubsection{One--Rim Flared Disk Model Fitting to SEDs Only}
\label{sed}

We first tune the model to fit the SEDs only, in order to evaluate how the predicted MIR visibilities compare with the data. The best--fit results are shown in Table~\ref{onerim_tab} and Figure \ref{onerim_fig1}. In addition to the best--fit parameters, Table~\ref{onerim_tab} includes the fractional MIR flux in each of the two disk components ($f^{MIR}_{rim}$ and $f^{MIR}_{disk}$), relative to the total MIR flux (star $+$ rim $+$ disk). We include no formal parameter errors, because our intent is not to determine precise parameter values, but to evaluate the validity of the main features of the model. The table also includes the reduced--$\chi^2$ values for the best--fit model compared to the SED and $V^2$ data; i.e. $\chi^2_{red} = \chi^2/(N - p)$, where $p = 3$ is the number of free parameters for the 1--rim model, and the number of data points $N$ is 10 for the visibility data, and of order 1000 (depending on the object) for the SED data.


For the 7 objects with observed silicate emission features, the SEDs are well reproduced. The NIR excess (``bump'') is due mostly to the rim as expected (and this validates the assumption made for the simple geometric model of the previous section). Most of the MIR flux arises in the surface layer of the disk, and reproduces the observed $10 \, \mu$m silicate peak well in most cases. 

What about the predicted MIR visibilities? For 3 of the objects: RY~Tau, MWC~758, and AB~Aur; which are the 3 most spatially resolved in the MIR, the visibility data are well reproduced. For all the other objects, the SED--best--fit model predicts MIR visibilities which are significantly lower than is observed; i.e. the data requires a much more ``compact'' MIR brightness. 

We conclude that, in general, the disk model when tuned to fit only the SEDs produces inadequate visibility predictions. This is an important observation, given that these models are in wide usage in the field, but the most common situation is the lack of spatially resolved data.


\subsubsection{One--Rim Flared Disk Model Fitting to SEDs and Visibilities}

We now use the same model to fit both the SED and MIR visibility data simultaneously. Since the SEDs have many more data points, we increase the weights of interferometer data accordingly (by the ratio of the number of data points). 

The results are shown in Figure \ref{onerim_fig2} and Table \ref{onerim_tab}, and can be summarized as follows: (1) For the 7 objects with observed silicate emission. (1a) A  solution that fits well both the SED and MIR visibility data now exists for SU~Aur, at a small cost in reduced agreement with the SED data. It can be seen in Table \ref{onerim_tab} that this is achieved by increasing the MIR flux contribution from the rim, resulting in a more compact source of MIR emission. (1b) We also note that for v1295~Aql, the fit to the visibility data is significantly improved, but at the expense of no longer fitting the SED well at all i.e. in this case, forcing more rim MIR emission results in greatly overshooting the NIR bump. (1c) In summary, a total of 4 the 7 objects with observed silicate emission are well fit by the model (the same 3 as in Section~\ref{sed} plus SU~Aur); for the other 3 the general feature remains that the MIR visibilities are lower than observed and a more compact MIR brightness is required. (2) For DG~Tau, which does not exhibit silicate features in the SED, we note that this model can match the SED and MIR visibility levels relatively well (except at the longer KIN wavelengths) perhaps indicating that the general features of the model have some applicability to this object. Interestingly, for the other 3 objects with no silicate feature in the SED (v1685~Cyg, MWC~1080 and v1057~Cyg) the general feature remains that the MIR visibilities are lower than observed and a more compact MIR brightness is required.

\subsubsection{Two--Rim Flared Disk Model Fitting to SEDs and Visibilities}

As shown above, the ``one--rim~+~disk'' model tends to underestimate the observed MIR visibilities, indicating that a significant fraction of the flux originates in a more compact source than predicted by this model. In fact, the required size scale for the MIR brightness is comparable to that of the inner rim, but this component alone cannot explain the observations because the relatively large dust grains required to survive direct exposure to the stellar radiation are not able to produce the required MIR flux.


Rather than attempting to tune this model, we explore here the viability of a more radical modification to the disk structure, motivated in part by the SED--modelling work of \citet{mcc2013}. The precise location and shape of the inner dust rim is determined by processes such as the settling of larger grains to the disk mid--plane and the dependence of dust sublimation temperature on the local gas density, dust grain size, and chemical composition, collectively leading to curved walls, which \citet{mcc2013} successfully model using a two--layer approximation. 

Here we implement the two--layer approximation using two distinct inner rims of different heights, but otherwise each modelled as in Section~\ref{sec:physmodel} (see Figure~\ref{fig: schematic disk}, compare with Figure~1 of \citet{mcc2013}). 
The second rim is located behind the first rim (further from the star), is taller than the first rim, and therefore still partially directly heated by the star. Thus smaller dust grains can survive in the second rim, which leads to the required compact MIR emission, compared to that arising in the extended disk behind it. 
The emission in the region between the two rims is difficult to predict due to possible rim--shadowing effects; thus for simplicity we assume no emission.
For the smaller dust grains of the second rim we adopt a size of 0.25~$\mu$m. We assume the dust composition of the two rims to be the same (described in \ref{sec:physmodel}). The 2--rim model therefore has two additional degrees of freedom: the scale parameter and temperature ($T_{rim2}$) of the second rim.  

The results are shown in Figure~\ref{fig: allfit_tworim} and Table~\ref{tworim_tab} (in the calculation of $\chi^2_{red}$, the number of model free parameters is now $p=5$). 
As expected, rim--2 (located at 1 to few AU) is cooler and contributes mainly to the MIR flux. In order to assess the {\em relative} quality of the 1~and 2--rim models, but taking into account the increased degrees of freedom for the 2--rim model, we use the Akaike Information Criterion (AIC): $\mbox{AIC} = 2 p + \chi^2_{red}$, where $p$ is the number of model free parameters. The $\mbox{AIC}$ still favors models with lower $\chi^2_{red}$, but penalizes for the increased degrees of freedom. Table~\ref{tworim_tab} shows $\Delta(\mbox{AIC}) = \mbox{AIC}_{2rim} - \mbox{AIC}_{1rim}$, a negative value favors the 2--rim model, which formally happens for 7 of the 11 objects when considering the fits to the SEDs, and for 5 of the 11 objects when considering the fits to the MIR visibilities.


We summarize the results as follows. For two of the objects, the 2--rim model still does not provide good fits, either because the MIR visibilities are not well fit (MWC~480) or because the SED is not well fit (v1295~Aql). Both objects have high $f^{MIR}_{rim2}$ and very low values of the disk flaring index (much lower than $\xi = 2/7$ for hydrostatic equilibrium), such that the flared disk has been essentially replaced by the second rim. For v1295~Aql, it may be that our model fails because contrary to our assumption the disk inclination is high (values in the literature range from $0 - 65$~deg, \citet{eis2004,ise2006}). Another possibility for this object is that the model is valid, but the dust properties in rim~2 need to be modified, given that as mentioned above this component dominates the MIR emission, and has the correct size scale, but fails mainly in that it significantly overpredicts the NIR fluxes. 

For all other cases the 2--rim model leads to improved results. For MWC~275, the MIR visibilities could not be fit at all by the 1--rim model, but the 2--rim model enables a good simultaneous fit to the SED and MIR visibilities. The same is true for MWC~758, but with a more modest $\chi^2_{V^2}$ improvement. For three other objects (SU~Aur, RY~Tau and AB~Aur) the 2--rim model maintains a similar fit to the MIR visibilities, but enables much improved fits to the SEDs, especially in the $\sim 5 - 12 \, \mu$m spectral region. 

\subsection{Additional comments on specific objects}

{\bf SU~Aur:} The disk mass is $\log(M_D)/M_\odot = -5.1^{+1.4}_{-0.8}$ \citep{ake2002}, relatively low compared to classic T~Tauri stars. Our 1--rim and 2--rim model solutions have the lowest disk flaring index, the disk near--flatness may be related to its low mass.

{\bf RY~Tau:} Formally, the 2--rim model is preferred. However, the second rim is located at 1.6~AU with temperature 1050~K; both similar to the typical size scale and temperature of the extended disk component in the one--rim disk model. This essentially indicates a degeneracy between the two models. 

{\bf MWC~758:} In this case the second rim and the extended disk contribute comparable MIR fluxes. Here again we obtain relatively low flaring indexes, in agreement with \citet{bes1999}. The second rim is located at 6.8~AU from the central star, much further than the 0.54~AU rim location in the 1--rim model. In other words, formally the 2--rim model replaces the inner $\sim 6.8$~AU of the extended disk with a narrow ring structure.

{\bf DG~Tau:} As noted above this is a very active object, with silicate emission that is variable on weeks timescales \citep{woo2004,bar2009} and sometimes appears in absorption \citep{sit2008}, perhaps indicating that a large amount of cool dust is lifted up above the disk surface and is causing self--absorption over the emission region \citep{tam2008}. 
At the epochs of our observations, we do not detect the $10 \, \mu$m silicate feature, while the coherent flux (orange solid line in Figure \ref{fig: geom}) 
suggests an absorption feature. Since KIN resolves the disk partially or fully, the coherent flux must come from regions smaller than the disk, implying that the lifted dust causing the absorption is located $\leq$ 1AU, a dynamical timescale consistent with the observed variation timescale of the silicate feature. 

\subsection{The spectral shape of the MIR visibilities}

We note that the MIR visibility data for our objects display a variety of spectral shapes: most are concave--up, but some are monotonically increasing (MWC~480, v1295~Aql) and MWC~275 is the only one with a concave--down shape (perhaps signaling a unique characteristic for this object). Our model is too simple to reproduce the shapes exactly, here we provide qualitatively arguments for how such differing spectral shapes can arise in a multi--component model for the emission. 

Consider a model where the MIR emission arises in two components -- as in the 2--rim model considered above, where rim~2 is a compact source of MIR emission relative to the extended disk behind it.  For the compact component, the visibilities will increase with wavelength as the angular resolution decreases at longer wavelengths. For the extended component, with a radial temperature profile such that the disk temperatures are lower at larger radii, the characteristic MIR size increases with wavelength and therefore the visibilities decrease; an effect that competes with the visibility increase due to lower angular resolution at longer wavelengths. The resulting shape will depend on the balance of these competing effects for the specific case of each object, as follows.

In the limiting case that the compact component is completely unresolved and the extended component is completely resolved, the MIR visibilities are equal to the fractional flux in the compact component, and a spectral concave--up shape will result if the MIR flux in the compact component is less peaked than that of the extended component. And viceversa for the concave--down spectral shape. In another limiting case, the compact component dominates the MIR fluxes, and the MIR visibilities increase monotonically with wavelength as a result of lower angular resolution. If on the other hand the extended component dominates the flux, either a concave--up or down shape can result depending which of the effects described above dominates.

\section{Summary and Conclusions}

We have measured the infrared visibilities and near--simultaneous SEDs of 11 young stellar objects, several of them spatially resolved at MIR wavelengths and long baselines for the first time. We use a simple geometrical model to provide basic information about the infrared brightness, namely the NIR and MIR size scales, independent of details of specific physical models. Further insight on the disk structure can be gained by studying how the characteristic sizes relate to the properties of the central star. The KIN MIR~sizes (measured as ($\overline{R}_{rim} + \overline{HWHM}_{Disk}$) of Section~\ref{sec:results})  appear better correlated with stellar luminosity than found by previous authors, although direct comparisons are complicated by the different models assumed. 

We test current disk paradigms for physical disk models in the form of a semi--analytical dust rim~$+$~disk model, and find that in several notable cases the model fails to reproduce the measured MIR visibilities and the SEDs simultaneously; with the data requiring relatively compact MIR emission (1 -- 7~AU). 

We explore the possibility that the MIR brightness is better modeled by taking into consideration the proposed layered morphology of the curved inner rim, which naturally leads to a series of inner--rims which containing different dust populations (grain sizes) \citep{mcc2013} and therefore contribute MIR emission on different size scales. We find that when implemented as a 2--rim approximation, the fits to the SEDs and MIR visibilities are significantly improved in most cases. 

We leave to future work extensions to the model which may alleviate the shortcomings of our 1 or 2--rim models, such as an exploration of the effects of varying the dust species or the inclusion of viscous heating processes. 

Instead of dust radial and scale height variations (layered disks), the 2--rim model could be mimicking structures due to forming planets (rings and gaps, \citet{menu2015}). However, the relatively high correlation found in our mid--infrared size vs. stellar luminosity relation favors layered disk morphologies, because of the higher stochasticity expected to be associated with early planet formation processes.

The detailed disk structure and brightness is likely to be complex, and to vary from object to object; emphasizing the need for theoretical progress driven by new observations. Spatially resolved MIR observations are a sensitive way to probe the disk vertical structure and time evolution, and our results highlight the fact that conventional smooth disk models developed to fit SEDs alone almost always fail to reproduce the MIR spatial scales. This is an important consideration in view of active on--going efforts to model circumstellar disks and the planet formation process within them. Improved baseline coverage and ultimately model--independent images of the inner disk at MIR wavelengths from the next generation VLTI/MATISSE instrument \citep{lop2014} 
or the proposed Planet Formation Imager \citep[PFI,][]{jdm2014} will be invaluable in offering a direct view of the $\sim 1 - 10$~AU planet formation region; much as the transformative knowledge gains now being delivered by ALMA observations of the cooler, more distant regions of pre--planetary disks \citep[e.g. the images for the HL Tau disk in][]{alma2015}. 

\appendix

\section{Validation of the semi--analytical flared disk model}

In order to verify our semi--analytical implementation of the flared disk model, we compare its predictions with the numerical radiative transfer benchmark models of \citet{pin2009}. The model parameters used for the benchmark comparison are shown in Table~\ref{benchmarkparams}. The parameters in our semi--analytical model are also set to best approximate the benchmark model; namely the flaring index is set to be 0.125 in order to match the disk scale height, and we suppress the dust inner rim emission, since this component is absent in the benchmark models.

Figure \ref{benchmark_comp} summarizes the comparisons between our semi--analytical model and the benchmark results for the case of the TORUS code in \citep{pin2009}. The left panel shows the SED comparison. As can be seen the semi--analytical model produces lower fluxes than TORUS from $2 - 20 \mu$m, perhaps due to scattering effects not included in our model \citep{dul2001}. The middle and right panels compare the spatial flux distribution, i.e. $\lambda F_{\lambda} \times R$ as a function of radius (T. Harries priv. comm., the right panel zooms in the inner disk in linear scale). The results are very similar, although the semi--analytical model produces more centrally peaked emission. These differences are not surprising given the very different detailed implementations of the model, and do not affect the conclusions in this paper. We conclude that for the purposes of this paper our semi--analytical implementation of the flared disk model has been validated.

\acknowledgments

The authors wish to acknowledge fruitful discussions with Nuria Calvet and Melissa McClure. Part of this work was performed while X. C. was a Visiting Graduate Student Research Fellow at the Infrared Processing and Analysis Center (IPAC), California Institute of Technology. The Keck Interferometer was funded by the National Aeronautics and Space Administration as part of its Exoplanet Exploration Program. Data presented herein were obtained at the W.M. Keck Observatory, which is operated as a scientific partnership among the California Institute of Technology, the University of California and the National Aeronautics and Space Administration. The Observatory was made possible by the generous financial support of the W.M. Keck Foundation. The authors wish to recognize and acknowledge the very significant cultural role and reverence that the summit of Mauna Kea has always had within the indigenous Hawaiian community.  We are most fortunate to have the opportunity to conduct observations from this mountain. Data presented in this paper were obtained at the Infrared Telescope Facility, which is operated by the University of Hawaii under contract NNH14CK55B with the National Aeronautics and Space Administration. We gratefully acknowledge support and participation in the IRTF/BASS observing runs by Daryl Kim, The Aerospace Corporation. This work has made use of services produced by the NASA Exoplanet Science Institute at the California Institute of Technology. M. S. was supported by NASA ADAP grant NNX09AC73G. R. W. R. was supported by the IR\&D program of The Aerospace Corporation.

{\it Facilities:} \facility{Keck: Interferometer}, \facility{IRTF: SpeX, BASS}

{}

\clearpage

\begin{figure} 
\begin{center} { \includegraphics[ trim = 30mm 30mm 35mm 35mm, scale = 0.75]{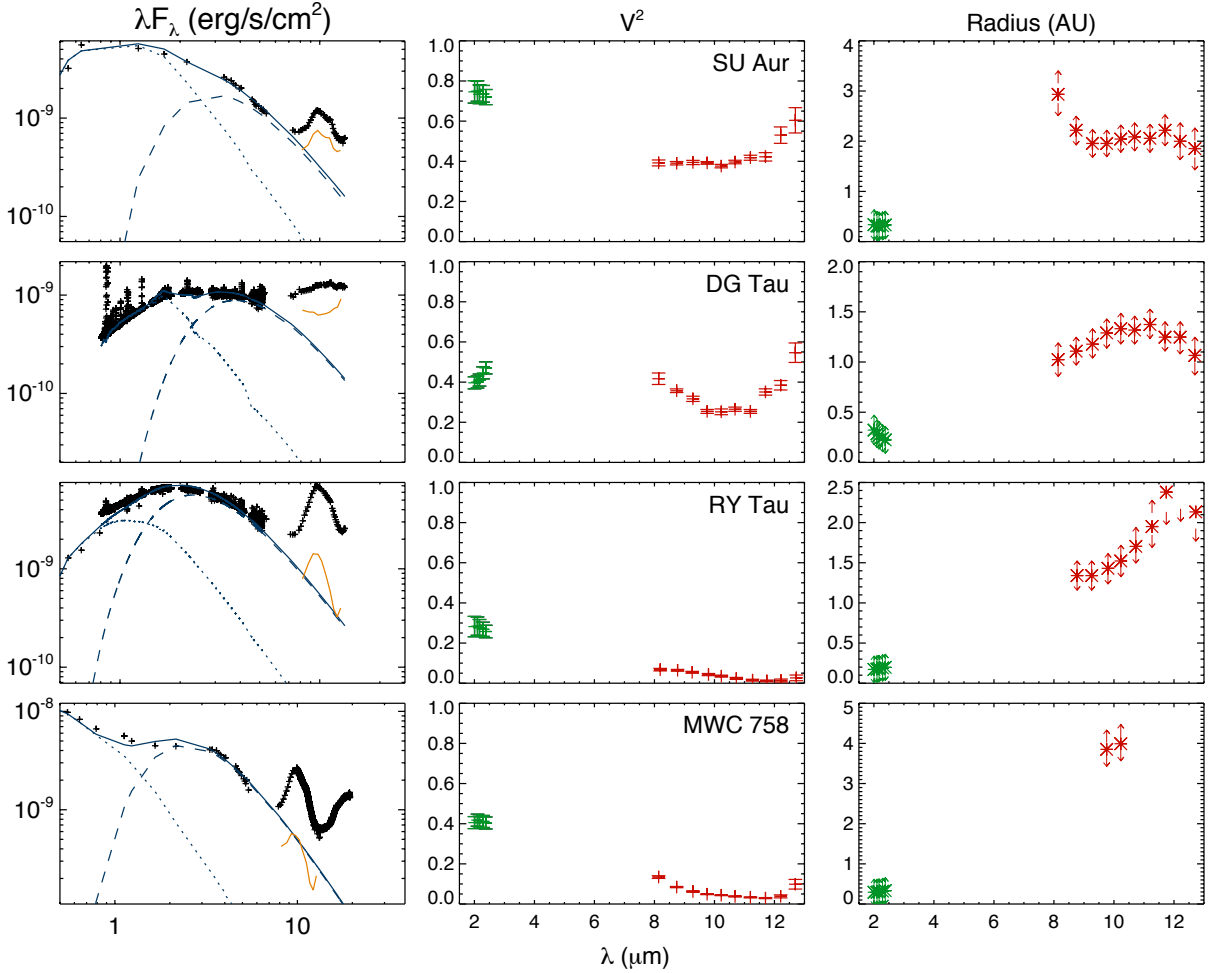} } 
\hphantom{.....} 
\caption{Interferometry and SED data, and results of fitting to the geometric model. 
The left panels show the SED data and models: stellar photosphere (blue dotted lines), blackbody ring representing the inner dust rim (blue dashed lines), star$+$rim (blue solid line), and coherent MIR flux (i.e. MIR flux times the visibility, orange lines). 
The middle panels show the NIR and MIR interferometer data (visibility modulus) for each of the NIR (green) and MIR (red) bandpasses.
The right panels show the best--fit characteristic radii as a function of wavelength in each of the bandpasses, i.e. at the NIR (green) wavelengths they are the best--fit radii of the ring representing the inner dust rim ($R_{rim}$), and at the MIR (red) wavelengths they are the best--fit $R_{rim}+\mbox{HWHM}_{Disk}$, where $\mbox{HWHM}_{Disk}$ is the half--width at half--maximum of the Gaussian brightness representing the extended disk (see text Section~\ref{sec:desc}).
The arrow symbols represent the upper and lower 1$\sigma$ range. In some cases 
the extended disk sizes are missing because no suitable solution exists (see text Section~\ref{sec:results}). 
\label{fig: geom} } 
\end{center} 
\end{figure}

\addtocounter{figure}{-1}
\begin{figure} 
\begin{center} { \includegraphics[ trim = 30mm 30mm 35mm 35mm, scale = 0.75]{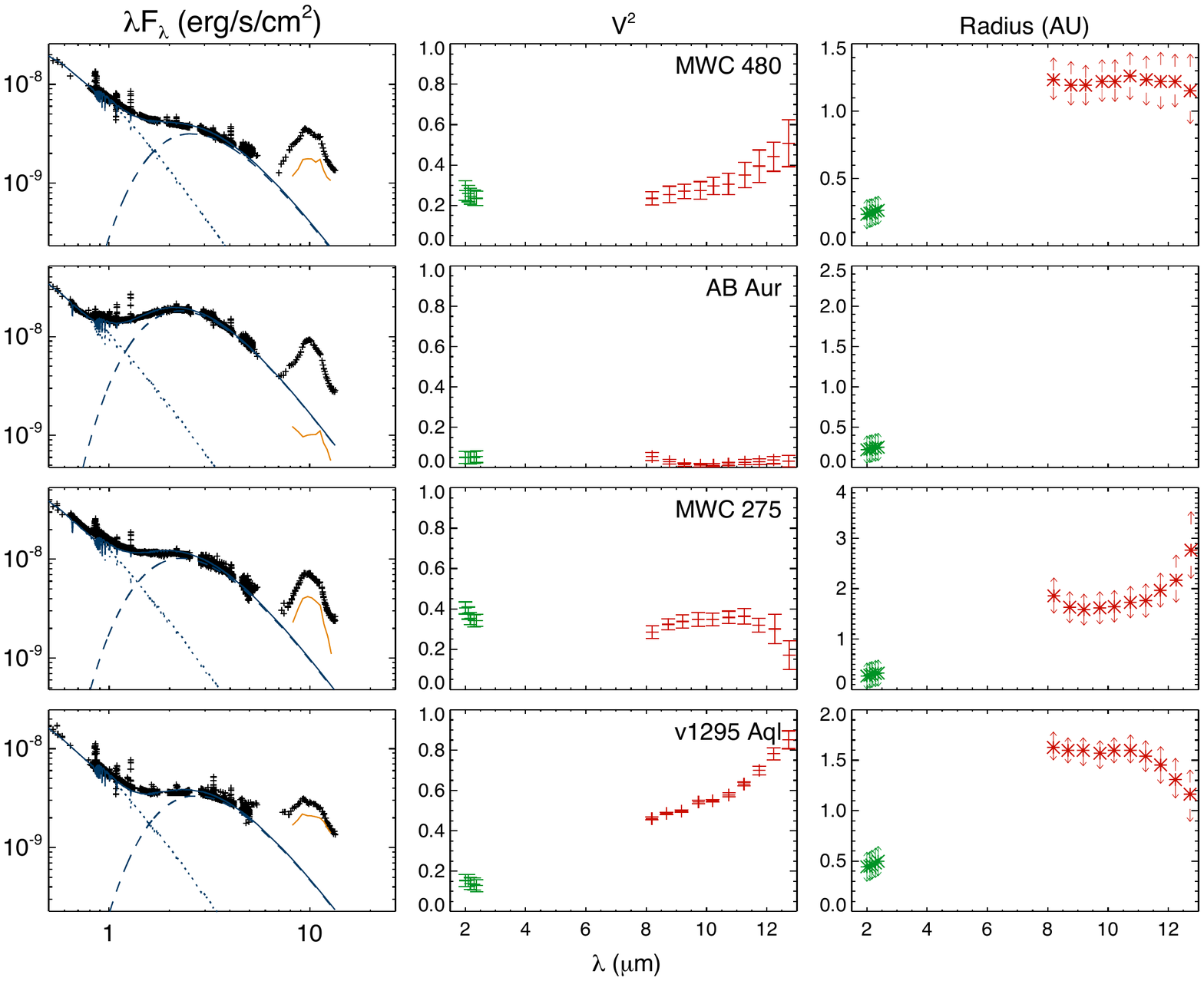} } 
\hphantom{.....} 
\caption{ Continued. 
\label{figure: simplegeo2} } 
\end{center} 
\end{figure} 

\addtocounter{figure}{-1}
\begin{figure} 
\begin{center} { \includegraphics[ trim = 30mm 30mm 35mm 35mm, scale = 0.75]{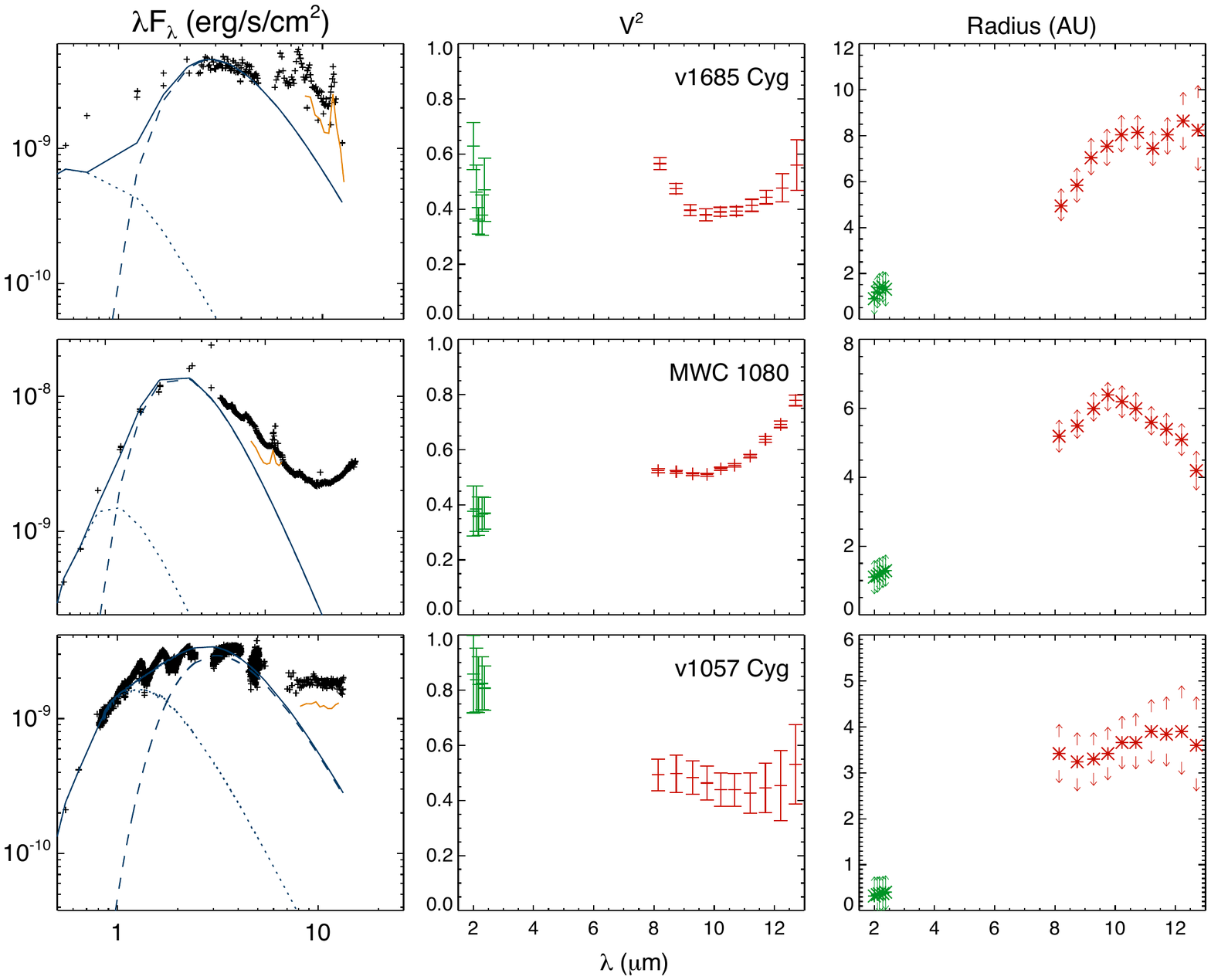} } 
\hphantom{.....} \caption{ Continued. 
\label{figure: simplegeo3} } 
\end{center}
\end{figure} 

\begin{figure} 
\begin{center} { \includegraphics[width = 6.5in]{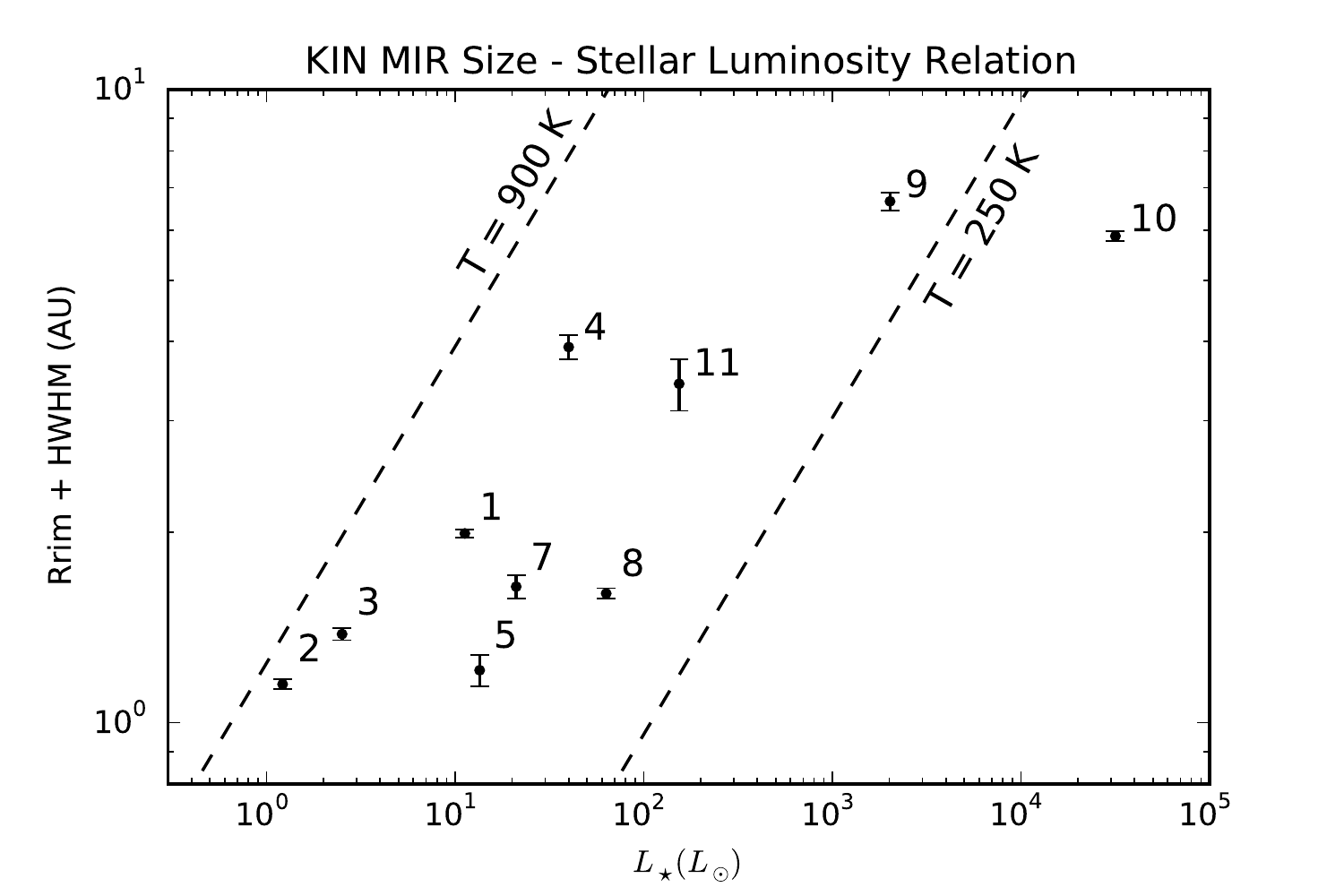} } 
\hphantom{.....} 
\caption{KIN MIR size -- stellar luminosity diagram. The index numbers identify each object as in Table~\ref{geom}. The dashed lines represent the equilibrium location of gray dust at the indicated temperatures.
\label{fig: size-l} } 
\end{center} 
\end{figure}

\begin{figure}
\begin{center}
{
\includegraphics[width = 6.5in,  trim = 30mm 30mm 35mm 30mm]{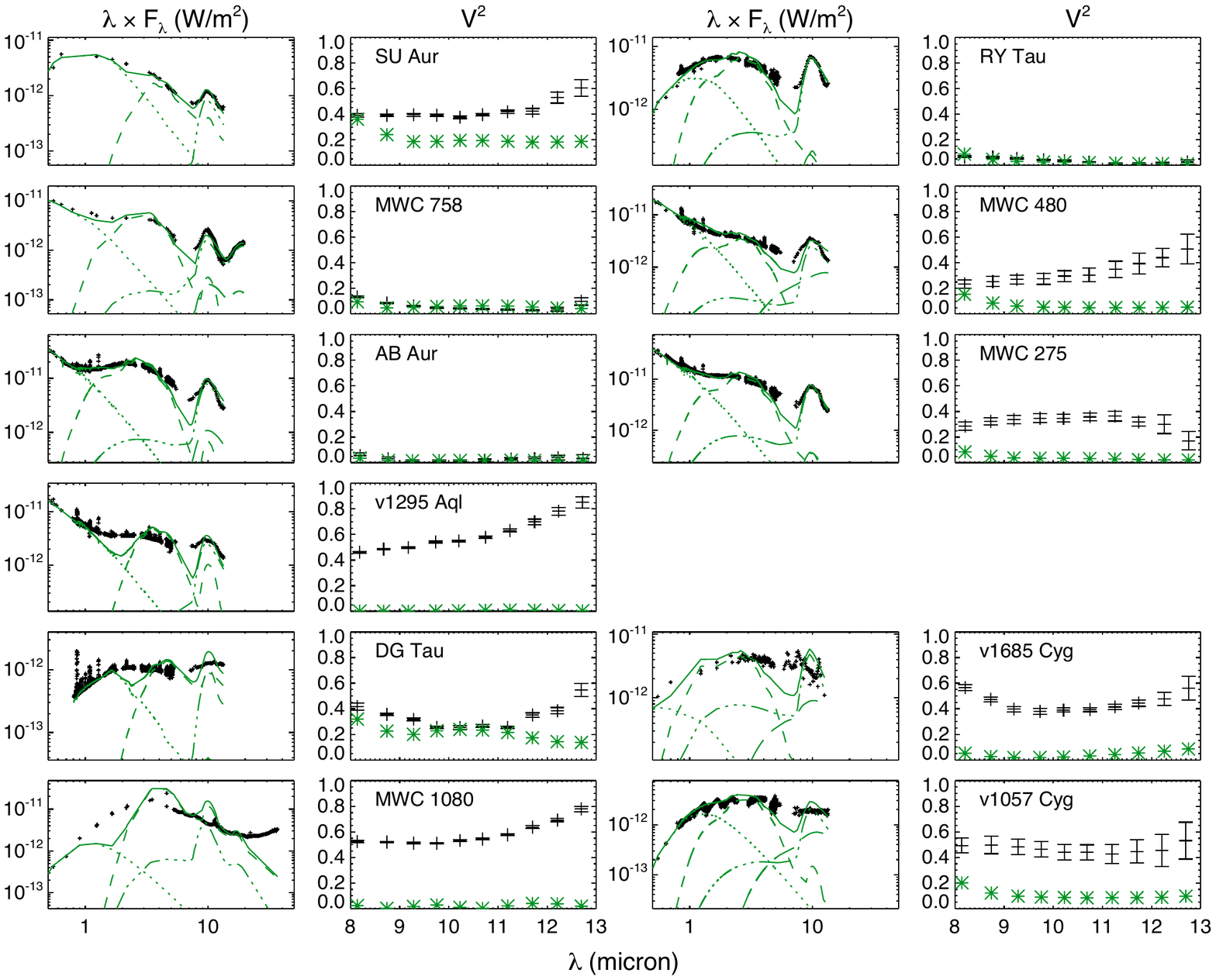}
}
\hphantom{.....}
\caption{ 
One rim disk model fitting to the SED data only. 
Data are shown as black symbols. The models are shown as green lines, as follows:
In the SED panels, the dotted line is the star, the short dashed line is the rim, 
the triple--dotted--dashed line is the surface layer, the long--dashed line is the interior layer, and the solid line is the total flux. The 4 objects in the bottom panels are the ones for which no silicate feature is observed in the SEDs, and are shown here for illustrative purposes and to evaluate how the models are able to reproduce the SED and visibility levels only. 
\label{onerim_fig1}
}
\end{center}
\end{figure}

\begin{figure}
\begin{center}
{
\includegraphics[width = 6.5in,  trim = 30mm 30mm 35mm 30mm]{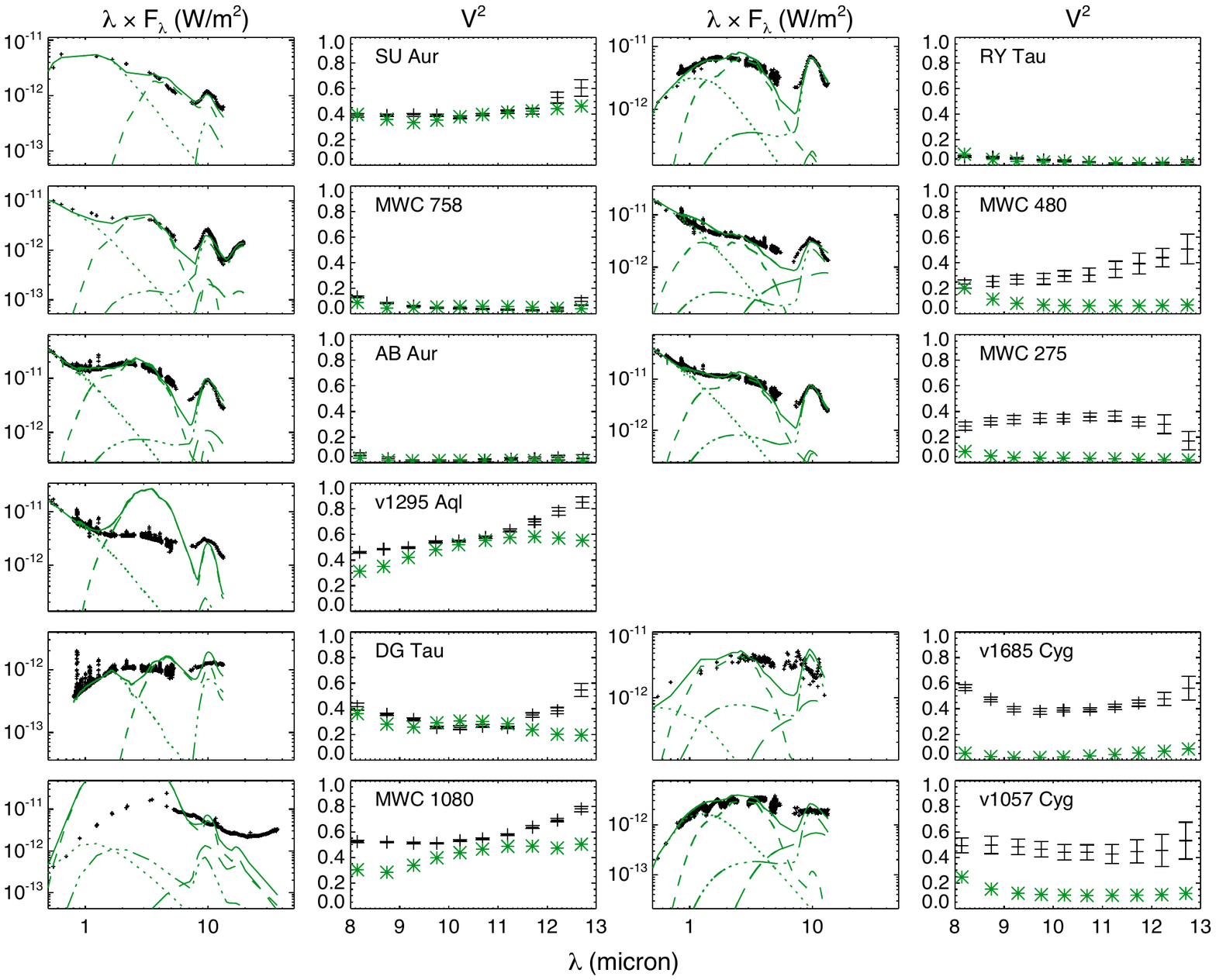}
}
\hphantom{.....}
\caption{ 
One rim disk model fitting to the SED and MIR interferometry data. The notations are the same as in Figure \ref{onerim_fig1}. The 4 objects in the bottom panels are the ones for which no silicate feature is observed in the SEDs, and are shown here for illustrative purposes and to evaluate how the models are able to reproduce the SED and visibility levels only. 
\label{onerim_fig2}
}
\end{center}
\end{figure}

\begin{figure}
\begin{center}
{
\includegraphics[width = 6.5in,  trim = 30mm 30mm 35mm 30mm]{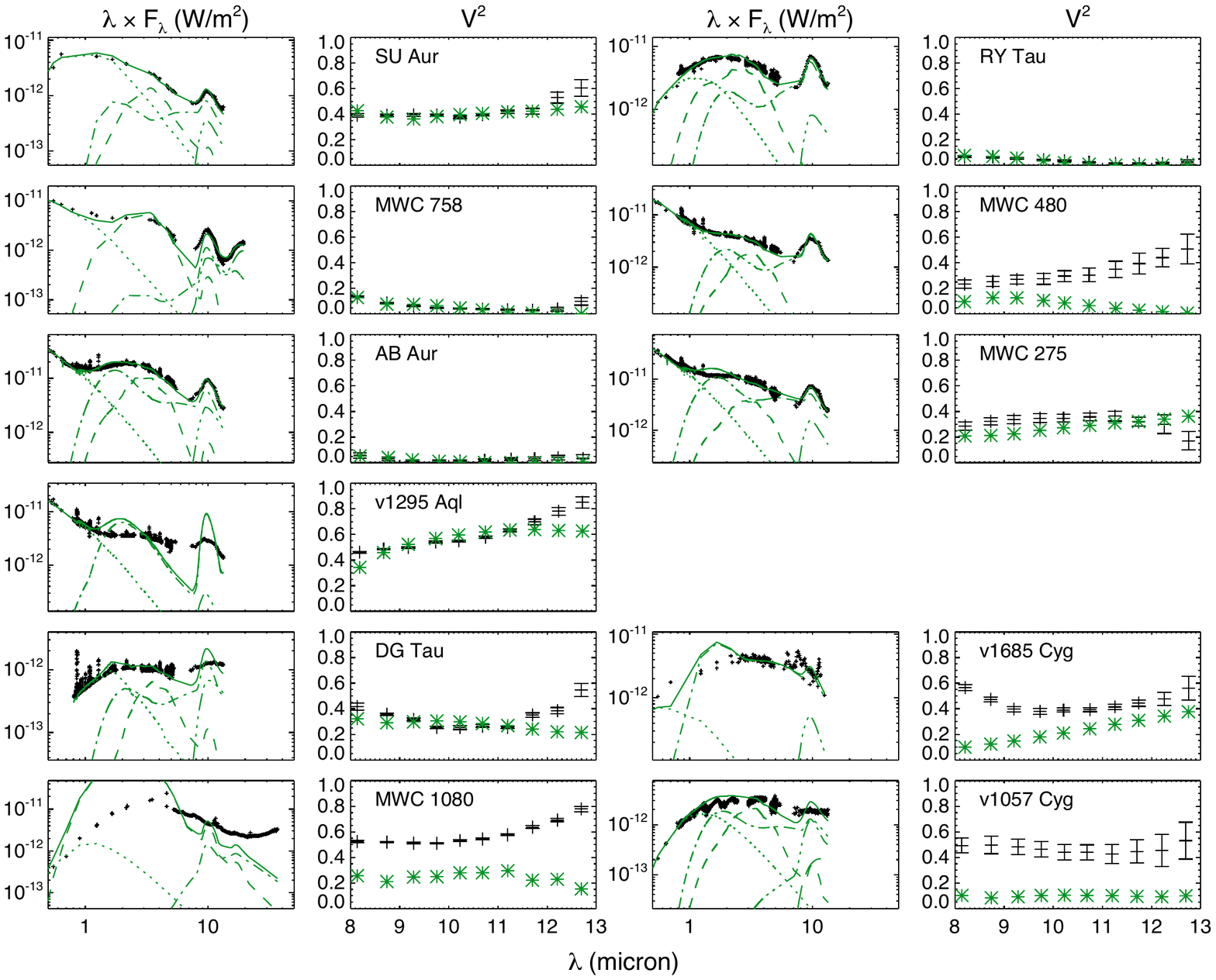}
}
\hphantom{.....}
\caption{ 
Two rim disk model fitting to the SED and interferometry data. The notations are the same as in Figure \ref{onerim_fig1}, with one addition: the dotted--dashed line represents the emission from the second rim. The 4 objects in the bottom panels are the ones for which no silicate feature is observed in the SEDs, and are shown here for illustrative purposes and to evaluate how the models are able to reproduce the SED and visibility levels only. 
\label{fig: allfit_tworim}
}
\end{center}
\end{figure}

\begin{figure} 
\begin{center} { \includegraphics[width = 6.5in,  trim = 30mm 145mm 35mm 20mm]{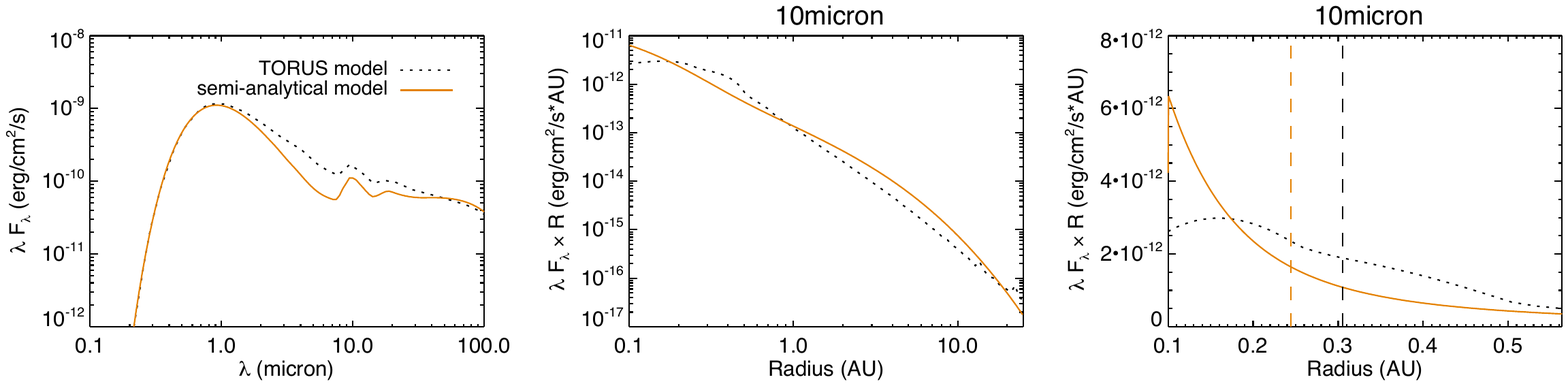} } 
\hphantom{.....} 
\caption{Benchmark comparison of our semi--analytical model with the radiative transfer TORUS model. The two vertical dashed lines in the right--most panel show the half--light radii. 
\label{benchmark_comp} } 
\end{center} 
\end{figure}

\clearpage 
\begin{deluxetable}{l c c c c c c c c c c }
\tabletypesize{\scriptsize}
\tablecaption{Target List}
\tablewidth{0pt}
\tablehead{
\colhead{Name} & \colhead{Hipparcos} & \colhead{Spectral } & \colhead{Mass} & \colhead{R.A.} & \colhead{Dec.}
& \colhead{V} & \colhead{K} & \colhead{N} & \colhead{d}  &  \colhead{Object} \\
 & & \colhead{Type} & \colhead{(M$_\sun$)} & & & \colhead{(mag)} & \colhead{(mag)} & \colhead{(Jy)} & \colhead{(pc)} & \colhead{Type}
}
\startdata
SU~Aur		& HIP22925	& G2		&2.63\tablenotemark{a}	& 04 55 59.385 	& +30 34 01.52		& 9.39	& 5.99	& 3.5  & 146\tablenotemark{e}	& TTS		\\
DG~Tau		& 			& K7		&0.91\tablenotemark{a}	& 04 27 04.698 	& +26 06 16.31		& 12.67	& 6.99	& 9.3  & 140\tablenotemark{f}	& TTS		\\
RY~Tau		& HIP20387	& K1		&2.27\tablenotemark{a}	& 04 21 57.410 	& +28 26 35.57		& 10.47	& 5.40	& 17.5 & 130\tablenotemark{g}	& TTS		\\
MWC~758		& HIP25793	& A3		&2.8\tablenotemark{c}	& 05 30 27.530 	& +25 19 57.08		& 8.27	& 5.80	& 4.6  & 279\tablenotemark{i}	& HAe		\\
MWC~480		& HIP23143	& A2		&3.08\tablenotemark{c}	& 04 58 46.265 	& +29 50 36.98		& 7.62	& 5.53	& 10.2 & 137\tablenotemark{e}	& HAe		\\
AB~Aur 		& HIP22910	& A1		&3.25\tablenotemark{a}	& 04 55 45.845 	& +30 33 04.29		& 7.06	& 4.23	& 27.2 & 140\tablenotemark{e}	& HBe		\\
MWC~275		& HIP87819	& A1		&2.3\tablenotemark{c}	& 17 56 21.288 	& -21 57 21.87		& 6.86	& 4.78	& 18.2 & 119\tablenotemark{e}	& HAe		\\
v1295~Aql	& HIP98719	& A0		&2.9\tablenotemark{d}	& 20 03 02.510 	& +05 44 16.67		& 7.73	& 5.86	& 7.2  & 290\tablenotemark{h}	& HAe		\\
v1685~Cyg	& HIP100289	& B3		&7.00\tablenotemark{a}	& 20 20 28.245 	& +41 21 51.56		&10.88	& 5.77	& 5.0  & 1000\tablenotemark{e}	& HBe		\\
MWC~1080	& HIP114995	& B0		&10.0\tablenotemark{b}	& 23 17 25.590 	& +60 50 43.62		& 11.86	& 4.83	& 22.2 & 1000\tablenotemark{h}	& HBe		\\
v1057~Cyg	&			& \nodata   &0.5\tablenotemark{j}	& 20 58 53.732 	& +44 15 28.54		&12.04	&6.23	& 5.7  & 600\tablenotemark{k}	&FUOR

\enddata
\tablenotetext{a}{\cite{kir11}}
\tablenotetext{b}{\cite{hoh10}}
\tablenotetext{c}{\cite{fol12}}
\tablenotetext{d}{\cite{alo09}}
\tablenotetext{e}{\cite{van07}}
\tablenotetext{f}{\cite{ken94}}
\tablenotetext{g}{\cite{kha09}}
\tablenotetext{h}{\cite{eis2004}}
\tablenotetext{i}{\cite{van98}}
\tablenotetext{j}{\cite{cla2005}}
\tablenotetext{k}{\cite{her03}}
\label{starlist}
\end{deluxetable}

\clearpage 
\begin{deluxetable}{l c c c c  c c }
\tabletypesize{\scriptsize}
\tablecaption{Disk properties from the literature}
\tablewidth{0pt}
\tablehead{
\colhead{Name} & \colhead{Inclination} & \colhead{P.A.} & \colhead{Mass} & \colhead{Outer radius} & \colhead{Observational} & \colhead{References} \\
& \colhead{(deg)} & \colhead{(deg)} & \colhead{(M$_\sun$)} & \colhead{(AU)} & \colhead{Technique} &
}
\startdata
SU~Aur		& $62^{+4}_{-8}$	& $127^{+8}_{-9}$ 	& 	$8\times10^{-6}$					& 70 - 240			& NIR/MIR interferometry 	& (b)\\
DG~Tau 		& $27 \pm 9$		& $120 \pm 24$		& 	$1-7\times10^{-4}$					& $72.3 \pm 4.0$	& mm interferometry		 	& (c) \\
RY~Tau 		& $66 \pm 2$		& $24 \pm 3$		& 	$3\times10^{-5} - 1.5\times10^{-4}$	& $70.5 \pm 3.9$	&  mm interferometry 		& (c)	\\
MWC~758		& $21 \pm 2$		& $65 \pm 7$		&	$1\times10^{-2}$					& $385\pm26$		& sub-mm/mm interferometry	& (j,h) \\
MWC~480		& $37 \pm 3$ 		& $143 \pm 5$ 		&	$6.1\times10^{-2}$					& 250				& mm interferometry			& (g) \\
AB~Aur 		& $21. \pm 0.5$		& $58.6 \pm 0.5$	& 	$9\times10^{-3}$					& $615^{+8}_{-3}$ 	& mm interferometry 		& (a) \\
MWC~275		& $48 \pm 2$		& $136 \pm 2$		&	$7\times10^{-4}$	 				& 200				& NIR/MIR/mm interferometry	& (i) \\
v1295~Aql 	& $0$				& $0$				&	$1.6\times10^{-4}$					& 100				& NIR/MIR interferometry 	& (f)	\\
v1685~Cyg 	& $41^{+3}_{-2}$	& $110^{+3}_{-4}$ 	&   $\le$0.133 							& \nodata 			& NIR interferometry 		& (d) \\
MWC~1080 	& $35^{+19}_{-16}$	& $54^{+13}_{-43}$	& 	$3.6\times10^{-3}$		        	& $77^{+23}_{-17}$ 	& NIR interferometry		& (d,e)	\\
v1057~Cyg 	& 30				& 177				&   $1\times10^{-1}$ 					& 200				& Spectroscopy/mm interferometry & (k,l,m)	\\
\enddata
\tablecomments{For v1295~Aql the disk inclination is very uncertain and we adopt a face--on geometry based on indications of low projected rotational velocity 
\citep{acke2004, pog2005} and interferometer data \citep{eis2004}. pionier paper in prep also gives low inc for v1295aql.
References: 
(a) \cite{cor05},
(b) \cite{ake2002},
(c) \cite{ise10a},
(d) \cite{eis2004},
(e) \cite{alo09},
(f) \cite{rag2012},
(g) \cite{ham06},
(h) \cite{cha08},
(i) \cite{tan2008b},
(j) \cite{ise10b},
(k) \cite{ken88},
(l) \cite{kos11},
(m) \cite{san01}
}
\label{disk properties}
\end{deluxetable}

\clearpage 
\begin{deluxetable}{l c c c c}
\tabletypesize{\scriptsize}
\tablecaption{Log of KIN and near--contemporaneous IRTF Observations}
\tablewidth{0pt}
\tablehead{
\colhead{Name} & \colhead{KIN} & \colhead{SpeX} & \colhead{BASS} & \colhead{KIN calibrators}
}
\startdata

SU~Aur		&UT09-21-2010		& \nodata		&UT10-23-2010 	& HD18449,HD52960 \\
DG~Tau		&UT09-21-2010		&UT11-25-2010	&UT10-23-2010 	& HD18449\\
RY~Tau		&UT10-26-2009		&UT11-25-2010	&UT10-23-2010 	& HD33463,HD39045 \\
MWC~758		&UT09-21-2010		& \nodata		&UT10-24-2010 	& HD18449,HD52960 \\
MWC~480		&UT10-26-2009		&UT12-01-2009	&UT11-29-2009 	& HD33463,HD39045 \\
AB~Aur 		&UT10-26-2009		&UT12-01-2009	&UT11-29-2009	& HD33463 , HD39045 \\
MWC~275		&UT07-07-2009		&UT07-08-2009 	&UT07-14-2009	& HD163197,HD169305,HD194193\\
			&					& \nodata		&UT07-16-2009	& HD214868,HD212496 \\
v1295~Aql	&UT06-02-2010		&UT08-22-2007	&UT10-22-2010	& HD203291 \\
v1685~Cyg	&UT07-07-2009		& \nodata		& \nodata 		& HD194093 \\
			&UT09-21-2010		& \nodata		& \nodata		& \nodata \\
MWC~1080	&UT09-21-2010		& \nodata		& \nodata 		& HD216946 \\
v1057~Cyg	&UT07-07-2009		&UT07-10-2009	&UT07-16-2009	& HD194193,HD209945,HD169305 \\
			&UT09-21-2010		& \nodata		&UT10-23-2010	& HD214868,HD212496\\
\enddata
\label{obslog}
\end{deluxetable}

\clearpage 
\begin{deluxetable}{l c c c c}
\tabletypesize{\scriptsize}
\tablecaption{Best--fit Stellar Parameters}
\tablewidth{0pt}
\tablehead{
\colhead{Name} & \colhead{$T_\star$} & \colhead{$R_\star$} & \colhead{E(B-V)}  & \colhead{$L_\star$} \\
& \colhead{(K)} & \colhead{(R$_\sun$)} & & \colhead{$L_\sun$}
}
\startdata

SU~Aur		& 5500		& 3.7		&0.40		& 11 \\
DG~Tau		& 4000		& 2.3		&0.80		& 1.2 \\
RY~Tau		& 5750		& 1.6		&0.71		& 2.5 \\
MWC~758		& 8250		& 3.1		&0.13		& 40 \\
MWC~480		& 8250		& 1.8		&0.02		& 13 \\
AB~Aur 	 	& 9000		& 2.1		&0.04		& 26 \\
MWC~275		& 8750		&2.0		&0.03		& 21 \\
v1295~Aql	& 9250		& 3.1		&0.08		& 63 \\
v1685~Cyg	& 22000		& 3.1		&0.80		& 2021 \\
MWC~1080	& 30000		& 6.6		&1.7		& 32000 \\
v1057~Cyg	& 6250		& 10.6		&1.35		& 154 \\
\enddata
\tablecomments{The stellar luminosities are derived from the best fit temperatures and radii.}
\label{stellarparam}
\end{deluxetable}

\clearpage 
\begin{deluxetable}{l l c c c }
\tabletypesize{\scriptsize}
\tablecaption{Geometrical model parameter results.}
\tablewidth{0pt}
\tablehead{
\colhead{Index} & \colhead{Name} & \colhead{$T_{rim}$} & \colhead{ $\overline{R}_{rim}$} & \colhead{$\overline{R}_{rim} + \overline{HWHM}_{Disk}$} \\
 &  & \colhead{(K)} & \colhead{(AU)} & \colhead{(AU)}
}
\startdata

1 & SU~Aur		& 1310$\pm$40	& 0.320$\pm$0.015	& 1.99$\pm$0.03	\\
2 & DG~Tau		& 1070$\pm$40	& 0.251$\pm$0.005	& 1.15$\pm$0.02	\\
3 & RY~Tau		& 1740$\pm$30	& 0.185$\pm$0.004	& 1.38$\pm$0.03	\\
4 & MWC~758		& 1550$\pm$20	& 0.309$\pm$0.005	& 3.92$\pm$0.17	\\
5 & MWC~480		& 1430$\pm$30	& 0.250$\pm$0.006	& 1.21$\pm$0.07	\\
6 & AB~Aur 	 	& 1630$\pm$10	& 0.236$\pm$0.008	& \nodata		\\
7 & MWC~275		& 1610$\pm$20	& 0.298$\pm$0.005	& 1.64$\pm$0.07	\\
8 & v1295~Aql	& 1370$\pm$20	& 0.465$\pm$0.010	& 1.60$\pm$0.03	\\
9 & v1685~Cyg	& 1400$\pm$10	& 1.28$\pm$0.05		& 6.66$\pm$0.22	\\
10 & MWC~1080	& 1580$\pm$20	& 1.21$\pm$0.05		& 5.87$\pm$0.11	\\
11 & v1057~Cyg	& 1330$\pm$30	& 0.377$\pm$0.061	& 3.43$\pm$0.32	\\

\enddata
\label{geom}
\end{deluxetable}

\clearpage 
\begin{deluxetable}{cccccccccccccccc}
\tabletypesize{\scriptsize}
\rotate
\tablecolumns{16}
\tablewidth{0pt}
\tablecaption{One--Rim Flared Disk Model Fitting Results}
\tablehead{
& \multicolumn{7}{c}{1 Rim -- fit to SED only} & \multicolumn{7}{c}{1 Rim -- fit to SED and $V^2$} \\
\cline{2-8} \cline{10-16} \\
\colhead{Name} & \colhead{$R_{rim1}$} & \colhead{$T_{rim1}$}  & \colhead{$f^{MIR}_{rim}$} & \colhead{$\xi$}  & \colhead{$f^{MIR}_{disk}$} & 
\colhead{$\chi^2_{red,SED}$} & \colhead{$\chi^2_{red,VIS}$} & & \colhead{$R_{rim1}$} & \colhead{$T_{rim1}$}   & \colhead{$f^{MIR}_{rim}$} & 
\colhead{$\xi$}   & \colhead{$f^{MIR}_{disk}$} & \colhead{$\chi^2_{red,SED}$} & \colhead{$\chi^2_{red,VIS}$} \\
& \colhead{(AU)} & \colhead{(K)} & & & & & & & \colhead{(AU)} & \colhead{(K)} & & & & &
}
\startdata
SU~Aur		& 0.94		& 950	&0.32	&0.14 	&0.65	&5.1	&253	& & 1.2  	& 850	&0.67	& 0.06	&0.29	&19		&7.3	\\
RY~Tau		& 0.12		& 2000	&0.03	&0.39	&0.97	&13		&5.7	& & 0.12	& 2000	&0.031	& 0.39	&0.96	&13		&5.7	\\
MWC~758		& 0.54		& 1550	&0.14	&0.17	&0.86	&8.2	&30		& & 0.54	& 1550	&0.13	& 0.17	&0.87	&8.4	&28		\\
MWC~480		& 0.15		& 2100	&0.024	&0.25	&0.97	&16		&23		& &  0.10	& 2500	&0.017	& 0.24	&0.98	&26		&19		\\
AB~Aur 		& 0.29		&1850 	&0.067	&0.25	&0.93	&11		&0.64	& &  0.29	& 1850	&0.067	& 0.25	&0.93	&10.8	&0.64	\\
MWC~275		& 0.25		&1850	&0.048	&0.21	&0.95	&9.4	&68		& &  0.24	& 1900	&0.046	& 0.21	&0.95	&9.7	&67		\\
v1295~Aql	& 2.2		& 950	&0.29	&0.18	&0.71	&67		&4162	& &  1.2	& 1250	&0.92	& 0.01	&0.08	&6415	&142	\\
\cutinhead{Objects with no observed silicate emission features}
DG~Tau		& 0.58		& 700	&0.50	&0.46	&0.49	&35		&59 	& &  0.58	& 700	&0.59	& 0.43	&0.40	&40		&29		\\
v1685~Cyg	& 1.2		& 2500	&0.019	&0.15	&0.98	&90		&334	& &  1.2	& 2500	&0.019	& 0.15	&0.98	&90		&334	\\
MWC~1080 	& 68		& 750	&0.78	&0.19	&0.22	&405	&5511	& &  4.5	& 2500	&0.74	& 0.01	&0.26	&955	&516	\\
v1057~Cyg	& 0.70		& 1850	&0.047	&0.12	&0.95	&18		&26		& &  0.55	& 2050	&0.041	& 0.11	&0.95	&25		&23		\\
\enddata
\tablecomments{Column descriptions (see Secs~\ref{sec:physmodel}). (1): Object name; (2,9): Rim radius; (3,10): Rim temperature; 
(4,6,11,13): Fractional MIR fluxes; (5,12): Disk flaring index.}
\label{onerim_tab}
\end{deluxetable}

\clearpage 
\begin{deluxetable}{ccccccccccccc}
\tabletypesize{\scriptsize}
\rotate
\tablecolumns{13}
\tablecaption{Two--Rim Flared Disk Model Fitting Results}
\tablewidth{0pt}
\tablehead{
\colhead{Name} &  \colhead{$R_{rim1}$} &  \colhead{$T_{rim1}$}   & \colhead{$f^{MIR}_{rim1}$} &  \colhead{$R_{rim2}$} &  \colhead{$T_{rim2}$} & \colhead{$f^{MIR}_{rim2}$} &
\colhead{$\xi$}  & \colhead{$f^{MIR}_{disk}$}   &  \colhead{$\chi^2_{red,SED}$} &  \colhead{$\chi^2_{red,VIS}$} & \multicolumn{2}{c}{$\Delta(\mbox{AIC})$} \\
 &  \colhead{(AU)} &  \colhead{(K)}   &  &  \colhead{(AU)} &  \colhead{(K)} & & 
 &   &   & & \colhead{SED} & \colhead{$V^2$}
}
\startdata
SU~Aur		& 0.45	& 1300	&0.096	& 1.2		& 1250	&0.59	& 0.07	&0.29	&5.4	&3.3		& -33 	& -3 \\
RY~Tau		& 0.10	& 2200	&0.098	& 1.6		& 1050	&0.83	& 0.21	&0.15	&4.8	&3.6		& -14 	& 6 \\
MWC~758		& 0.54	& 1550	&0.14	& 6.8		& 800	&0.48	& 0.16	&0.38	&7.2	&9.3		& 24 	& -38 \\
MWC~480		& 0.44	& 1350	&0.031	& 2.3		& 1050	&0.95	& 0.01	&0.013	&3.2	&20			& -78 	& 84 \\
AB~Aur 	 	& 0.57	& 1400	&0.074	& 7.5		& 1350	&0.58	& 0.17	&0.34	&3.7	&1.9		& -14 	& 13 \\
MWC~275		& 0.46	& 1450	&0.034	& 1.3		& 1400	&0.72	& 0.10	&0.24	&10		&6.1		& 41 	& -219 \\
v1295~Aql	& 0.39	& 1950	&0.0001	& 1.1		& 1750	&0.96	& 0.02	&0.04	&110	&43			& -2500 & -224 \\
\cutinhead{Objects with no observed silicate emission features}
DG~Tau		& 0.25	& 1050	&0.25	& 0.65		& 1000	&0.56	& 0.43	&0.38	&11		&23			& -72 	& 68 \\
v1685~Cyg	& 2.5	& 1800	&$2x10^{-7}$	& 7.5	& 1600	&0.84	& 0.04	&0.16	&41		&129	& -32 	& -304 \\
MWC~1080	& 4.5	& 2400	&0.72	& 310	& 500	&0.28	& 0.01	&0.0	&619	&1707			& 1132 	& 11592 \\
v1057~Cyg	& 1.3	& 1450	&0.054	& 3.3	& 1400	&0.39	& 0.1	&0.55	&17		&25				& 36   	& 108 \\
\enddata
\tablecomments{Column descriptions (see Secs~\ref{sec:physmodel}). (1): Object name; (2,5): Rim radius; (3,6): Rim temperature; 
(4,7,9): Fractional MIR fluxes; (8): Disk flaring index.}
\label{tworim_tab}
\end{deluxetable}

\clearpage 
\begin{deluxetable}{l c  }
\tabletypesize{\scriptsize}
\tablecaption{Stellar and disk parameters for the benchmark code comparisons.}
\tablewidth{0pt}
\tablehead{
\colhead{Description} & \colhead{Value or expression} 
}
\startdata

Stellar temperature				& 4000 K									\\
Stellar radius					& 2~R$_\sun$									\\
Disk vertical profile				&  $\rho(r,z) = \rho_0(r)e^{-z^2/2h(r)^2}$			\\
Disk scale height				& $h(r) = (10AU)(r/100AU)^{1.125}$					\\
Disk surface density				& $\Sigma(r) = \Sigma_0(r/100AU)^{-1.5}$			\\
Total disk mass					& $3 \times 10^{-5} M_\sun$						\\
Disk inner radius				& 0.1~AU									\\
Disk outer radius				& 400~AU			\\
Dust grain size					& $1 \mu$m			\\
Dust grain density				& 3.5 $g/cm^3$									\\
Dust grain material				& silicates								\\

\enddata
\tablecomments{Parameters are the same as in the benchmark paper \citet{pin2009}.}
\label{benchmarkparams}
\end{deluxetable}


\begin{thebibliography}{}

\bibitem[ALMA Partnership et al.(2015)]{alma2015} ALMA Partnership, Brogan, C.~L., P{\'e}rez, L.~M., et al.\ 2015, \apjl, 808, L3 

\bibitem[Acke 
\& van den Ancker(2004)]{acke2004} Acke, B., \& van den Ancker, M.~E.\ 2004, \aap, 426, 151 

\bibitem[Alonso-Albi et 
al.(2009)]{alo09} Alonso-Albi, T., Fuente, A., Bachiller, R., et al.\ 2009, \aap, 497, 117 

\bibitem[Akeson et al.(2002)]{ake2002} Akeson, R.~L., Ciardi, 
D.~R., van Belle, G.~T., \& Creech-Eakman, M.~J.\ 2002, \apj, 566, 1124 

\bibitem[van Boekel et al.(2004)]{boek2004} van Boekel, R., Min, 
M., Leinert, C., et al.\ 2004, \nat, 432, 479 

\bibitem[van den Ancker et 
al.(1998)]{van98} van den Ancker, M.~E., de Winter, D., \& Tjin A Djie, H.~R.~E.\ 1998, \aap, 330, 145 

\bibitem[Bary et al.(2009)]{bar2009} Bary, J.~S., Leisenring, 
J.~M., \& Skrutskie, M.~F.\ 2009, \apjl, 706, L168 

\bibitem[Benisty et 
al.(2010)]{ben2010} Benisty, M., Tatulli, E., M{\'e}nard, F., \& Swain, M.~R.\ 2010, \aap, 511, AA75 

\bibitem[Beskrovnaya et al.(1999)]{bes1999} Beskrovnaya, N.~G., Pogodin, M.~A., Miroshnichenko, A.~S., et al.\ 1999, \aap, 343, 163 

\bibitem[Bodenheimer 
\& Lin(2002)]{bod2002} Bodenheimer, P., \& Lin, D.~N.~C.\ 2002, Annual Review of Earth and Planetary Sciences, 30, 113

\bibitem[van Boekel et 
al.(2005)]{boek2005} van Boekel, R., Dullemond, C.~P., \& Dominik, C.\ 2005, \aap, 441, 563 

\bibitem[Blum 
\& Wurm(2008)]{blum2008} Blum, J., \& Wurm, G.\ 2008, \araa, 46, 21 

\bibitem[Chapillon et 
al.(2008)]{cha08} Chapillon, E., Guilloteau, S., Dutrey, A., \& Pi{\'e}tu, V.\ 2008, \aap, 488, 565 

\bibitem[Chiang \& Goldreich(1997)]{chi1997} Chiang, E.~I., \& Goldreich, P.\ 1997, \apj, 490, 368 

\bibitem[Clarke et al.(2005)]{cla2005} Clarke, C., Lodato, G., 
Melnikov, S.~Y., \& Ibrahimov, M.~A.\ 2005, \mnras, 361, 942 

\bibitem[Colavita et al.(2009)]{col2009} Colavita, M.~M., 
Serabyn, E., Millan-Gabet, R., et al.\ 2009, \pasp, 121, 1120 

\bibitem[Colavita et al.(2010)]{col2010} Colavita, M.~M., 
Serabyn, E., Ragland, S., Millan-Gabet, R., 
\& Akeson, R.~L.\ 2010, \procspie, 7734, 77340T 

\bibitem[Colavita et al.(2013)]{col2013} Colavita, M.~M., 
Wizinowich, P.~L., Akeson, R.~L., et al.\ 2013, \pasp, 125, 1226 

\bibitem[Corder et al.(2005)]{cor05} Corder, S., Eisner, J., 
\& Sargent, A.\ 2005, \apjl, 622, L133 

\bibitem[Cushing et al.(2004)]{cus2004} Cushing, M.~C., Vacca, 
W.~D., \& Rayner, J.~T.\ 2004, \pasp, 116, 362 

\bibitem[D'Alessio et al.(2004)]{dal2004} D'Alessio, P., 
Calvet, N., Hartmann, L., Muzerolle, J., 
\& Sitko, M.\ 2004, Star Formation at High Angular Resolution, 221, 403 

\bibitem[Dorschner et al.(1995)]{dor1995} Dorschner, J., Begemann, B., Henning, T., Jaeger, C., \& Mutschke, H.\ 1995, \aap, 300, 503 

\bibitem[Dullemond et al.(2001)]{dul2001} Dullemond, C.~P., 
Dominik, C., \& Natta, A.\ 2001, \apj, 560, 957 

\bibitem[Dullemond 
\& Monnier(2010)]{dul2010} Dullemond, C.~P., \& Monnier, J.~D.\ 2010, \araa, 48, 205 

\bibitem[Eisner et al.(2004)]{eis2004} Eisner, J.~A., Lane, 
B.~F., Hillenbrand, L.~A., Akeson, R.~L., 
\& Sargent, A.~I.\ 2004, \apj, 613, 1049 

\bibitem[Fedele et 
al.(2008)]{fed2008} Fedele, D., van den Ancker, M.~E., Acke, B., et al.\ 2008, \aap, 491, 809 

\bibitem[di Folco et 
al.(2009)]{folco2009} di Folco, E., Dutrey, A., Chesneau, O., et al.\ 2009, \aap, 500, 1065 

\bibitem[Folsom et al.(2012)]{fol12} Folsom, C.~P., Bagnulo, 
S., Wade, G.~A., et al.\ 2012, \mnras, 422, 2072 

\bibitem[Gab{\'a}nyi  et al.(2013)]{gab2013} Gab{\'a}nyi , 
K.~{\'E}., Mosoni, L., Juh{\'a}sz, A., et al.\ 2013, Astronomische 
Nachrichten, 334, 912 

\bibitem[Hackwell et al.(1990)]{hac1990} Hackwell, J.~A., 
Warren, D.~W., Chatelain, M.~A., Dotan, Y., 
\& Li, P.~H.\ 1990, \procspie, 1235, 171 

\bibitem[Hamidouche et al.(2006)]{ham06} Hamidouche, M., 
Looney, L.~W., \& Mundy, L.~G.\ 2006, \apj, 651, 321 

\bibitem[Herbig et al.(2003)]{her03} Herbig, G.~H., Petrov, 
P.~P., \& Duemmler, R.\ 2003, \apj, 595, 384 

\bibitem[Hinz et al.(2001)]{hinz2001} Hinz, P.~M., Hoffmann, 
W.~F., \& Hora, J.~L.\ 2001, \apjl, 561, L131 

\bibitem[Hohle et al.(2010)]{hoh10} Hohle, M.~M., 
Neuh{\"a}user, R., 
\& Schutz, B.~F.\ 2010, Astronomische Nachrichten, 331, 349 

\bibitem[Isella \& Natta(2005)]{ise2005} Isella, A., \& Natta, A.\ 2005, \aap, 438, 899 

\bibitem[Isella et al.(2006)]{ise2006} Isella, A., Testi, L., \& Natta, A.\ 2006, \aap, 451, 951 

\bibitem[Isella et al.(2010)]{ise10a} Isella, A., Carpenter, 
J.~M., \& Sargent, A.~I.\ 2010, \apj, 714, 1746 

\bibitem[Isella et al.(2010)]{ise10b} Isella, A., Natta, A., 
Wilner, D., Carpenter, J.~M., \& Testi, L.\ 2010, \apj, 725, 1735 

\bibitem[Jaeger et al.(1994)]{jae1994} Jaeger, C., Mutschke, H., Begemann, B., Dorschner, J., \& Henning, T.\ 1994, \aap, 292, 641 

\bibitem[Kharchenko 
\& Roeser(2009)]{kha09} Kharchenko, N.~V., \& Roeser, S.\ 2009, VizieR Online Data Catalog, 1280, 0 

\bibitem[Kenyon et al.(1988)]{ken88} Kenyon, S.~J., Hartmann, 
L., \& Hewett, R.\ 1988, \apj, 325, 231 

\bibitem[Kenyon et al.(1994)]{ken94} Kenyon, S.~J., 
Dobrzycka, D., \& Hartmann, L.\ 1994, \aj, 108, 1872 

\bibitem[Kirk 
\& Myers(2011)]{kir11} Kirk, H., \& Myers, P.~C.\ 2011, \apj, 727, 64 

\bibitem[K{\'o}sp{\'a}l(2011)]{kos11} K{\'o}sp{\'a}l, {\'A}.\ 2011, \aap, 535, AA125 

\bibitem[Kraus et al.(2008)]{kraus2008} Kraus, S., Preibisch, T., 
\& Ohnaka, K.\ 2008, \apj, 676, 490 

\bibitem[Kurucz(1979)]{kur1979} Kurucz, R.~L.\ 1979, \apjs, 40, 1 

\bibitem[Laor \& Draine(1993)]{lao1993} Laor, A., \& Draine, B.~T.\ 1993, \apj, 402, 441 

\bibitem[Leinert et 
al.(2004)]{lein2004} Leinert, C., van Boekel, R., Waters, L.~B.~F.~M., et al.\ 2004, \aap, 423, 537 

\bibitem[van Leeuwen(2007)]{van07} van Leeuwen, F.\ 2007, 
Astrophysics and Space Science Library, 350,  

\bibitem[Liu et al.(2007)]{liu2007} Liu, W.~M., Hinz, P.~M., 
Meyer, M.~R., et al.\ 2007, \apj, 658, 1164 

\bibitem[Liu et al.(2005)]{liu2005} Liu, W.~M., Hinz, P.~M., 
Hoffmann, W.~F., et al.\ 2005, \apjl, 618, L133 

\bibitem[Lopez et al.(2014)]{lop2014} Lopez, B., Lagarde, S., 
Jaffe, W., et al.\ 2014, \procspie, 9146, 91460M 

\bibitem[Maaskant et 
al.(2013)]{maa2013} Maaskant, K.~M., Honda, M., Waters, L.~B.~F.~M., et al.\ 2013, \aap, 555, A64 

\bibitem[McClure et al.(2013)]{mcc2013} McClure, M.~K., 
D'Alessio, P., Calvet, N., et al.\ 2013, \apj, 775, 114 

\bibitem[Menu et al.(2015)]{menu2015} Menu, J., van Boekel, R., Henning, T., et al.\ 2015, \aap, 581, A107 

\bibitem[Millan-Gabet et al.(2011)]{mill11} Millan-Gabet, R., 
Serabyn, E., Mennesson, B., et al.\ 2011, \apj, 734, 67 

\bibitem[Mathis et al.(1977)]{mat1977} Mathis, J.~S., Rumpl, 
W., \& Nordsieck, K.~H.\ 1977, \apj, 217, 425 

\bibitem[Mennesson et al.(2014)]{men2014} Mennesson, B., 
Millan-Gabet, R., Serabyn, E., et al.\ 2014, \apj, 797, 119 

\bibitem[Millan-Gabet et al.(2011)]{mil2011} Millan-Gabet, R., 
Serabyn, E., Mennesson, B., et al.\ 2011, \apj, 734, 67 

\bibitem[Monnier et al.(2014)]{jdm2014} Monnier, J.~D., Kraus, 
S., Buscher, D., et al.\ 2014, \procspie, 9146, 914610 

\bibitem[Monnier et al.(2009)]{jdm2009} Monnier, J.~D., 
Tuthill, P.~G., Ireland, M., et al.\ 2009, \apj, 700, 491 

\bibitem[Monnier \& Millan-Gabet(2002)]{m&m2002} Monnier, J.~D., \& Millan-Gabet, R.\ 2002, \apj, 579, 694 

\bibitem[Oudmaijer et 
al.(2001)]{oud2001} Oudmaijer, R.~D., Palacios, J., Eiroa, C., et al.\ 2001, \aap, 379, 564 

\bibitem[Pinte et 
al.(2009)]{pin2009} Pinte, C., Harries, T.~J., Min, M., et al.\ 2009, \aap, 498, 967 

\bibitem[Pogodin et 
al.(2005)]{pog2005} Pogodin, M.~A., Franco, G.~A.~P., \& Lopes, D.~F.\ 2005, \aap, 438, 239 

\bibitem[Ragland et al.(2012)]{rag2012} Ragland, S., Ohnaka, 
K., Hillenbrand, L., et al.\ 2012, \apj, 746, 126 

\bibitem[Ratzka et 
al.(2009)]{ratz2009} Ratzka, T., Schegerer, A.~A., Leinert, C., et al.\ 2009, \aap, 502, 623 

\bibitem[Rayner et al.(2003)]{ray2003} Rayner, J.~T., Toomey, 
D.~W., Onaka, P.~M., et al.\ 2003, \pasp, 115, 362 

\bibitem[Sandell 
\& Weintraub(2001)]{san01} Sandell, G., \& Weintraub, D.~A.\ 2001, \apjs, 134, 115 

\bibitem[Sargent et al.(2009)]{sar2009} Sargent, B.~A., 
Forrest, W.~J., Tayrien, C., et al.\ 2009, \apjs, 182, 477 

\bibitem[Schegerer et 
al.(2013)]{sche2013} Schegerer, A.~A., Ratzka, T., Schuller, P.~A., et al.\ 2013, \aap, 555, AA103 

\bibitem[Schegerer et 
al.(2009)]{sche2009} Schegerer, A.~A., Wolf, S., Hummel, C.~A., Quanz, S.~P., \& Richichi, A.\ 2009, \aap, 502, 367 

\bibitem[Schegerer et 
al.(2008)]{sche2008} Schegerer, A.~A., Wolf, S., Ratzka, T., \& Leinert, C.\ 2008, \aap, 478, 779 

\bibitem[Serabyn et al.(2012)]{ser2012} Serabyn, E., Mennesson, 
B., Colavita, M.~M., Koresko, C., \& Kuchner, M.~J.\ 2012, \apj, 748, 55 

\bibitem[Sitko et al.(2008)]{sit2008} Sitko, M.~L., Carpenter, 
W.~J., Kimes, R.~L., et al.\ 2008, \apj, 678, 1070 

\bibitem[Snow 
\& Witt(1995)]{sno1995} Snow, T.~P., \& Witt, A.~N.\ 1995, Science, 270, 1455 

\bibitem[Tambovtseva 
\& Grinin(2008)]{tam2008} Tambovtseva, L.~V., \& Grinin, V.~P.\ 2008, Astronomy Letters, 34, 231 

\bibitem[Tannirkulam et al.(2008a)]{tan2008a} Tannirkulam, A., 
Monnier, J.~D., Harries, T.~J., et al.\ 2008a, \apj, 689, 513 

\bibitem[Tannirkulam et al.(2008b)]{tan2008b} Tannirkulam, A., 
Monnier, J.~D., Millan-Gabet, R., et al.\ 2008b, \apjl, 677, L51 

\bibitem[Vacca et al.(2003)]{vac2003} Vacca, W.~D., Cushing, 
M.~C., \& Rayner, J.~T.\ 2003, \pasp, 115, 389 

\bibitem[Williams 
\& Cieza(2011)]{wil2011} Williams, J.~P., \& Cieza, L.~A.\ 2011, \araa, 49, 67 

\bibitem[de Winter et 
al.(2001)]{win2001} de Winter, D., van den Ancker, M.~E., Maira, A., et al.\ 2001, \aap, 380, 609 

\bibitem[Woodward et al.(2004)]{woo2004} Woodward, C.~E., 
Wooden, D.~H., Harker, D.~E., et al.\ 2004, Debris Disks and the Formation 
of Planets, 324, 224 

\end{thebibliography}
\end{document}